\documentclass[runningheads]{svmult}

\usepackage{makeidx}   
\usepackage{graphicx}  
\usepackage{subeqnar}  
\usepackage{multicol}  
\usepackage{physprbb}  
\makeindex             

\newcommand{\be}{\begin{equation}}
\newcommand{\ee}{\end{equation}}
\newcommand{\bea}{\begin{eqnarray}}
\newcommand{\eea}{\end{eqnarray}}
\newcommand{\eq}[1]{Eq.~(\ref{#1})}
\newcommand{\Dslash}{D\!\!\!\!/}

\begin{document}
\title*{Hadron Properties with FLIC Fermions}
%
%
%
%
\titlerunning{FLIC Fermions}
%
\author{J.M.~Zanotti\inst{1,2}
\and D.B.~Leinweber\inst{1}
\and W.~Melnitchouk\inst{3}
\and A.G.~Williams\inst{1} 
\and J.B.~Zhang\inst{1}}
\authorrunning{J.M.~Zanotti et al.}
%
%
\institute{Department of Physics and Mathematical Physics and\\
        Special Research Centre for the
        Subatomic Structure of Matter,                          \\
        University of Adelaide, 5005, Australia
\and John von Neumann-Institut f\"ur Computing
  NIC, \\
  Deutsches Elektronen-Synchrotron DESY, D-15738 Zeuthen, Germany
\and Jefferson Lab, 12000 Jefferson Avenue,
        Newport News, VA 23606 }

\maketitle              

 \vspace{-7cm}
 \null \hfill ADP-04-09/T591 \\
 \null \hfill DESY-04-076 \\
 \null \hfill JLAB-04-226 \\
 \vspace{-20pt}
 \vspace{-20pt}
 \vspace{-20pt}
 \vspace{7cm}

\begin{abstract}
The Fat-Link Irrelevant Clover (FLIC) fermion action provides a new
form of nonperturbative ${\mathcal O}(a)$-improvement in lattice
fermion actions offering near continuum results at finite lattice
spacing.  It provides computationally inexpensive access to the light
quark mass regime of QCD where chiral nonanalytic behaviour associated
with Goldstone bosons is revealed.  The motivation and formulation of
FLIC fermions, its excellent scaling properties and its low-lying
hadron mass phenomenology are presented.
\end{abstract}

\section{Introduction}

The origin of the masses of light hadrons represents one of the most
fundamental challenges to QCD.
Despite the universal acceptance of QCD as the basis from which to derive
hadronic properties, there has been slow progress in understanding the
generation of hadron mass from first principles.
Solving the problem of the hadronic mass spectrum would allow considerable
improvement in our understanding of the nonperturbative nature of QCD.
The only available method at present to derive hadron masses directly
from QCD is a numerical calculation on the lattice.

The high computational cost required to perform accurate lattice
calculations at small lattice spacings has led to
increased interest in quark action improvement.
In this article we present results of simulations of the
spectrum of light mesons and baryons using an ${\cal O}(a)$ improved
fermion action \cite{FATJAMES,Proc,LQMpaper,Leinweber:2002bw}.
%
%
In particular, we will start with the standard clover action and replace
the links in the irrelevant operators with APE smeared
\cite{Falcioni:1985ei,Albanese:1987ds}, or fat links.
We shall refer to this action as the Fat-Link Irrelevant Clover (FLIC)
action.
Although the idea of using fat links only in the irrelevant operators of
the fermion action was developed here independently, suggestions have
appeared previously \cite{neub}.

In Section~\ref{QuarkAction}, we provide the reader with some
background information on lattice fermion actions. In particular, we
start with the basic lattice discretisation of the derivative
appearing in the continuum Dirac action, followed by improvements
suggested first by Wilson \cite{Wilson} and then by Sheikholeslami and
Wolhert \cite{Sheikholeslami:1985ij}.  Section~\ref{FLinks} contains
the procedure for creating the FLIC fermion action while in
Section~\ref{simulations} we describe the gauge configurations used in
our lattice simulations.
The results of an investigation of the scaling of this action at
finite lattice spacing are presented in Section \ref{discussion}.  In
Section~\ref{exceptionalSection} we investigate the problem of
exceptional configurations by performing simulations of hadron masses
at light quark masses corresponding to $m_\pi / m_\rho = 0.35$.
Section~\ref{splitSection} discusses the evidence for enhancement in
octet-decuplet mass splittings as one approaches the chiral limit in
the quenched approximation and finally in Section \ref{FLICconclusion}
we summarise the results.

\section{The Lattice Quark Action}
\label{QuarkAction}
\subsection{The Naive Fermion Action}

In Euclidean space-time, the continuum Dirac action 
is written as
\be
\bar{\psi}\, (\Dslash + m)\, \psi \, ,
\label{Daction}
\ee
where the covariant derivative is defined as $D_\mu = \partial_\mu +
i\, g\, A_\mu$.  Wilson \cite{Wilson} discretised the continuum Dirac
action by replacing the derivative with a symmetrised finite
difference.  Gauge links are included to not only encode the gluon
field, $A_{\mu}$, but to also maintain gauge invariance
\be
\bar{\psi}\, \Dslash \, \psi =
  \frac{1}{2a}\, \bar{\psi}(x)\, \sum_{\mu}\, \gamma_{\mu}\,
   \bigg[ U_{\mu}(x)\, \psi(x+a \hat{\mu})-
U^{\dagger}_{\mu}(x - a \hat{\mu})\, \psi(x - a \hat{\mu} ) \bigg]\, .
\label{discreteD}
\ee
Our notation uses the Pauli (Sakurai) representation of the Dirac
$\gamma$-matrices defined in Appendix~B of Sakurai \cite{Sakurai}.  In
particular, the $\gamma$-matrices are hermitian, $\gamma_{\mu} =
\gamma_{\mu}^{\dagger}$ and $\sigma_{\mu\nu} = [\gamma_{\mu},\,
\gamma_{\nu}]/(2i)$ such that $\gamma_\mu \, \gamma_\nu = \delta_{\mu
\nu} + i\, \sigma_{\mu \nu}$.
The gauge link variables at space-time position $x$ are defined as
\be
U_{\mu}(x) = {\cal P} \exp{\Big\{i\, g\, \int^{a}_0 \, A_{\mu}(x+\lambda
  \hat{\mu}) \, d\lambda \Big \}}\, .
\label{links}
\ee
Here the operator ${\cal P}$ path-orders the $A_{\mu}$'s along the
integration path, $a$ is the lattice spacing, and $g$ is the coupling
constant.

The continuum Dirac action is recovered in the limit $a\rightarrow 0$
by  Taylor expanding the $U_{\mu}$ and $\psi(x+a\hat{\mu})$ in
powers of the lattice spacing $a$. Keeping only the leading term in
$a$ (and for ease of notation we write
$a\hat{\mu}\rightarrow\hat{\mu}$), Eq.~(\ref{discreteD}) becomes 
\begin{eqnarray}
\frac{1}{2a}\, \bar{\psi}(x)\, \gamma_{\mu}\, \bigl[ &(& 1+i\, a\, g\,
  A_{\mu}(x+\frac{\hat{\mu}}{2}) + \ldots)\, (\psi(x) + a\, \psi'(x) +
  \ldots) - 
\nonumber \\
&(& 1 - i\, a\, g\, A_{\mu}(x-\frac{\hat{\mu}}{2}) + \ldots) (\psi(x) -
a\, \psi'(x) + \ldots )\bigr] \nonumber \\
= \bar{\psi}(x)\, \gamma_{\mu}\, (\partial_{\mu} &+& {\cal O}(a^2)) \psi(x)
+ i\, g\, \bar{\psi}(x)\, \gamma_{\mu}
\bigl[ A_{\mu} + {\cal O}(a^2)
\bigr]\psi(x), \label{TaylorEx}
\end{eqnarray}
which is the kinetic part of the standard continuum Dirac action in
Euclidean space-time to ${\cal O}(a^2)$.
Hence we arrive at the simplest (``naive'') lattice fermion action,
\begin{eqnarray} 
S_N &=& m_q \sum_x \bar{\psi}(x)\, \psi(x) + \nonumber \\
&& \frac{1}{2a} \sum_{x,\mu} \bar{\psi}(x)\, \gamma_{\mu}\, 
   \bigg[ U_{\mu}(x)\, \psi(x+\hat{\mu}) -
     U^{\dagger}_{\mu}(x-\hat{\mu})\, 
   \psi(x-\hat{\mu}) \bigg] \nonumber \\
&\equiv& \sum_x \bar{\psi}(x)\, M^N_{xy}[U]\, \psi(y),
\label{naiveQaction}
\end{eqnarray}
where the interaction matrix $M^N$ is 
\be
M_{i,j}^N [U] = m_q\, \delta_{ij} + \frac{1}{2a} \sum_{\mu} [
\gamma_{\mu}\, U_{i,\mu}\, \delta_{i,j-\mu} - \gamma_{\mu}\, 
U^{\dagger}_{i-\mu,\mu}\, \delta_{i,j+\mu}] .
\ee
The Taylor expansion in \eq{TaylorEx} shows that the naive fermion
action of \eq{naiveQaction} has ${\cal O}(a^2)$ errors. It also
preserves chiral symmetry.  However, in the continuum limit it gives
rise to $2^d = 16$ flavours of quark rather than one.  This is the famous
doubling problem and is easily demonstrated by considering the inverse
of the free field propagator (obtained by taking the fourier transform
of the action with all $U_{\mu} = 1$)
\be
S^{-1}(p) = m_q + \frac{i}{a} \sum_{\mu} \gamma_{\mu} \sin p_{\mu} a
\, ,
\ee
which has 16 zeros within the Brillouin cell in the limit $m_q
\rightarrow 0$. eg, $p_\mu = (0,0,0,0),\, (\pi /a ,0,0,0),\, (\pi /a
,\pi /a ,0,0)$, etc. Consequently, this action is phenomenologically
not acceptable. 

There are two approaches commonly used to remove these doublers. The
first involves adding operators to the quark action which scale with
the lattice spacing and thus vanish in the continuum limit. These
operators are chosen to drive the doublers to high energies and hence
are suppressed. This technique for improving fermion actions proceeds
via the improvement scheme proposed by Symanzik \cite{Symanzik:1983dc} and is
discussed in more detail in the following sections. The second method
for removing doublers involves ``staggering'' the quark degrees of
freedom on the lattice.  This procedure exploits the fact that the
naive fermion action has a much larger symmetry group, $U_V (4)
\otimes U_A (4)$, to reduce the doubling problem from $2^d = 16
\rightarrow 16/4$ while maintaining a remnant chiral symmetry. This
approach is not used in this article so the details of the action will
not be discussed here. Details of the derivation of staggered fermions
can be found in most texts (eg.  \cite{Rothe,Gupta}).

\subsection{Wilson Fermions}

In order to avoid the doubling problem, Wilson \cite{Wilson}
originally introduced an irrelevant (energy) dimension-five operator
(the ``Wilson term'') to the standard naive lattice fermion action
(\eq{naiveQaction}), which explicitly breaks chiral symmetry at 
${\cal O}(a)$.
\be
S_{\rm W} = \bar{\psi}(x) \left[ \sum_{\mu} \left( \gamma_{\mu}\, \nabla_{\mu}
    - \frac{1}{2}\, r\, a\, \Delta_{\mu} \right) + m \right] \psi(x)\ ,
\label{WilsonQaction}
\ee
where $r$ is the ``Wilson coefficient,''
\be
\nabla_{\mu}\psi(x) = \frac{1}{2a} [ U_{\mu}(x)\, \psi(x+\hat{\mu}) -
U^{\dagger}_{\mu}(x-\hat{\mu})\, \psi(x-\hat{\mu})] \, ,
\ee
and
\be
\Delta_{\mu}\psi(x) = \frac{1}{a^2} [ U_{\mu}(x)\, \psi(x+\hat{\mu}) +
U^{\dagger}_{\mu}(x-\hat{\mu})\, \psi(x-\hat{\mu}) -2\, \psi(x)].
\ee
The Wilson action in \eq{WilsonQaction} has no doublers for $r>0$ as
the Wilson term gives the extra fifteen species at $p_{\mu} = \pi/a$ a
mass proportional to $r/a$. Also, if $r=1$ then the Wilson action has
no ghost branches in its dispersion relation (see, for example,
Ref.~\cite{Alford:1996nx}).  In terms of link variables, $U_{\mu}(x)$, the
Wilson action can be written
\begin{eqnarray} 
S_W &=& \left(m_q + \frac{4r}{a}\right) \sum_x \bar{\psi}(x)\, \psi(x)
+ \frac{1}{2a} \sum_{x,\mu} \bar{\psi}(x) 
\bigg[  (\gamma_{\mu} - r)\, U_{\mu}(x)\, \psi(x+\hat{\mu}) \nonumber \\
&& \qquad \qquad - (\gamma_{\mu} +r)\,
  U^{\dagger}_{\mu}(x-\hat{\mu})\, \psi(x-\hat{\mu}) \bigg] \, , \\
&\equiv& \sum_{x,y} \bar{\psi}^L_x\,  M_{xy}^W\, \psi^L_y \, ,
\end{eqnarray}
where the interaction matrix for the Wilson action, $M^W$, is usually
written
\be
M_{xy}^W [U]\, a  = \delta_{xy} - \kappa \sum_{\mu} \Bigl [
(r-\gamma_{\mu})\, U_{x,\mu}\, \delta_{x,y-\mu} + (r+\gamma_{\mu})\,
U^{\dagger}_{x-\mu,\mu}\, \delta_{x,y+\mu} \Bigr] \, ,
\ee
with a field renormalisation
\begin{eqnarray}
\kappa &=& 1/(2\, m_q\, a + 8\, r) \, ,\nonumber \\
\psi^L &=& \psi / \sqrt{2\, \kappa} \, .
\end{eqnarray}
We take the standard value $r=1$ and the quark mass is given by
\be
m_q a = \frac{1}{2}\left( \frac{1}{\kappa} - \frac{1}{\kappa_c}
\right) \, .
\ee
In the free theory the critical value of kappa, $\kappa_c$, where the
quark mass vanishes, is $1/8r$. In the interacting theory, $\kappa_c$
is renormalised away from $1/8r$.  The quark mass has both
multiplicative and additive renormalisations due to the explicit
breaking of chiral symmetry by the Wilson term.

In the continuum limit, one finds
\be
S_W = \int d^4 x\, \bar{\psi}(x)\, 
\left(\Dslash + m - \frac{a\, r}{2}\sum_{\mu}D_{\mu}^2 \right )\, \psi (x) +
{\cal O}(a) \, .
\ee
By lifting the mass of the unwanted doublers with a second derivative,
${\cal O}(a)$ discretisation errors have been introduced into the
fermion matrix.  In contrast, the Wilson gauge action \cite{Wilson}
has only ${\cal O}(a^2)$ errors.  Hence there is enormous interest in
applying Symanzik's improvement program \cite{Symanzik:1983dc} to the
fermion action to remove ${\cal O}(a)$ errors by adding additional
higher dimension terms.

\subsection{Improving The Fermion Action}
\label{IMPfermions}

The addition of the Wilson term to the fermion action introduces large
${\cal O}(a)$ errors which means that in order to extrapolate reliably
to the continuum limit, simulations must be performed on fine lattices,
which are therefore very computationally expensive.  The scaling
properties of the Wilson action at finite $a$ can be improved by
introducing any number of irrelevant operators of increasing dimension
whose contributions vanish in the continuum limit.

The first attempt at removing these ${\cal O}(a)$ errors was by Hamber
and Wu \cite{Hamber:1983qa}
who added a two link term to the Wilson action
\begin{eqnarray}
S_{HW} = S_W &+& \kappa \sum_{x,\mu} \left [ \bar{\psi}^L(x)\,
\left ( -\frac{1}{4}\, r + \frac{1}{8}\, \gamma_\mu \right )
U_{\mu}(x)\, U_{\mu}(x+\hat{\mu})\, \psi^L(x+2\hat{\mu}) \right .
\nonumber \\ 
&+& \left . \bar{\psi}^L(x+2\hat{\mu})\, \left ( -\frac{1}{4}\, r -
\frac{1}{8}\, \gamma_\mu
\right ) U^{\dagger}_{\mu}(x+\hat{\mu})\, U^{\dagger}_{\mu}(x)\, 
\psi^L(x) \right ].
\label{HamberWu}
\end{eqnarray}
The removal of the classical ${\cal O}(a)$ terms is easily observed
through a Taylor expansion.  While this action also removes ${\cal
  O}(a^2)$ errors at tree-level, it has only received a small amount
of interest due to the computational expense in evaluating the double
hopping term. Calculations that have been done with this action show
that it works well at coarse lattice spacings and has the added bonus
that it has an improved dispersion relation \cite{Lee:1997bq}.

A more systematic approach \cite{Symanzik:1983dc} to ${\cal O}(a)$
improvement of the lattice fermion action in general
\cite{Sheikholeslami:1985ij} is to consider all possible gauge
invariant, local dimension-five operators, respecting the symmetries
of QCD
\begin{eqnarray}
{\cal O}_1 &=& -\frac{i\, g\, a\, C_{SW}\,
  r}{4}\bar{\psi}\, \sigma_{\mu\nu}\, F_{\mu\nu}\, \psi ,\nonumber \\ 
{\cal O}_2 &=& c_2\, a\, \left\{ \bar{\psi}\, D_{\mu}\, D_{\mu}\, \psi +
  \bar{\psi}\, \overleftarrow{D}_{\mu}\, \overleftarrow{D}_{\mu}\,
  \psi\right\} , \nonumber \\ 
{\cal O}_3 &=& \frac{b_g\, a\, m_q}{2}\, {\rm tr} \left\{ F_{\mu\nu}\,
  F_{\mu\nu} \right\} 
, \label{D5operators} \\
{\cal O}_4 &=& c_4\, m_q\, \left \{\bar{\psi}\, \gamma_{\mu}\, D_{\mu}\, \psi -
  \bar{\psi}\, \overleftarrow{D}_{\mu}\, \gamma_{\mu}\, \psi \right \} ,
  \nonumber \\
{\cal O}_5 &=& -b_m\, a\, m_q^2\, \bar{\psi}\, \psi . \nonumber
\end{eqnarray}
Note that the operator $\Dslash\, \Dslash$ is linearly related to
  ${\cal O}_1$ and ${\cal O}_2$ as
\be
\Dslash\, \Dslash = D_\mu \, D_\mu - \frac{g}{2}\, \sigma_{\mu \nu} \,
F_{\mu \nu}  \, .
\ee
Operator ${\cal O}_1$ is a new local operator in the lattice fermion
action and must be included.  On the other hand, ${\cal O}_3$ and
${\cal O}_5$ of \eq{D5operators} act to simply renormalise the
coefficients of existing terms in the lattice action, removing ${\cal
O}(a\, m_q)$ terms from the relation between bare and renormalised
quantities.  For example, the renormalisation of the quark mass $m_q
\rightarrow m_q\, (1 - b_m\, a\, m_q)$ incorporates ${\cal O}_5$.
Similarly, ${\cal O}_3$ introduces a mass dependence in the gauge
coupling $g^2 \rightarrow g^2\, (1 - b_g\, a\, m_q)$ such that the
lattice spacing remains constant for constant $g$ as $m_q$ is varied
\cite{Dawson:1997gp}.  For the quenched approximation, $b_g = 0$.

The key observation to efficient ${\cal O}(a)$ improvement is that the
${\cal O}(a)$ improvement afforded by two-link terms of the fermion
action \cite{Heatlie:1990kg} may be incorporated to ${\cal O}(a)$ into
the standard Wilson fermion action complemented by ${\cal O}_1$ though
the following transformation of the fermion fields
\begin{eqnarray}
\psi &\rightarrow& \psi' = (1+b_q\, r\, a\, m_q)\, (1 - c_q\, r\, a\,
\Dslash )\, \psi \, , 
\nonumber \\
\bar{\psi} &\rightarrow& \bar\psi' = (1+b_q\, r\, a\, m_q)\,
\bar{\psi}\, (1 + c_q\, r\, a\, \overleftarrow{\Dslash} ) \, ,
\label{fieldrot}
\end{eqnarray}
where $\psi'$ represents the physical fermion field recovered in the
continuum limit, while $\psi$ is the lattice fermion field used in the
numerical simulations.  The Jacobian of the transformation is 1 to
${\cal O}(a)$ \cite{Sheikholeslami:1985ij}.  At tree-level, $b_q = c_q
= 1/4$ correctly incorporates the ${\cal O}(a)$ corrections of ${\cal
O}_2$ and ${\cal O}_4$ into the fermion action.  Note that this field
transformation renormalises the fermion mass of the simulations
%
%
\be
m \rightarrow m ( 1 + \frac{1}{2}\, r\, a\, m) \, .
\ee

For the spectral quantities investigated herein, it is sufficient to
work with the lattice fermion-field operators alone.  Here the fermion
operators, act only as interpolators between the QCD vacuum and the
state of interest and do not affect the eigenstates of the QCD
Hamiltonian.  However, for matrix elements of fermion operators,
hadron decay constants, or off-shell quantities such as the quark
propagator, it is important to take this field redefinition into
account to connect the lattice field operators to the continuum field
operators incorporating important ${\cal O}(a)$ contributions.

In summary, ${\cal O}_1$, the ``clover'' term, is the only
dimension-five operator explicitly required to complement the Wilson
action to obtain ${\cal O}(a)$ improvement.  This particular action is
known as the Sheikholeslami-Wohlert fermion
\cite{Sheikholeslami:1985ij} action
\be
S_{SW} = S_{W} - \frac{i\, g\, a\, C_{SW}\, r}{4}\
         \bar{\psi}(x)\, \sigma_{\mu\nu}\, F_{\mu\nu}\, \psi(x)\ ,
\label{clover}
\ee
where $C_{SW}$  is the clover coefficient which can be tuned to
remove ${\cal O}(a)$ artifacts to all orders in the gauge coupling
constant $g$.
\begin{eqnarray}
C_{SW} &=&
\left\{
\begin{tabular}{l}
        1  \, \mbox{\rm{\ \ \ at tree-level\, ,}}       \\
        $1/u_0^3$\, \mbox{\rm{mean-field improved\, ,}}
\end{tabular}
\right.
\end{eqnarray}
with $u_0$ the tadpole improvement factor correcting for the
quantum renormalisation of the operators (see definition in
Section~\ref{simulations}).
Nonperturbative (NP) ${\cal O}(a)$ improvement \cite{Luscher:1996sc}
uses the axial Ward identity to tune $C_{SW}$ and remove all ${\cal
O}(a)$ artifacts provided one simultaneously improves the coupling
$g^2$, the quark mass $m_q$, and the currents \cite{Luscher:1996sc}.
The advantage of the clover action is that it is local and is only a
small overhead on Wilson fermion simulations.  Further details
of the improvement the clover action provides at finite lattice
spacing is given in Section~\ref{discussion}.

\begin{figure}[t]
\begin{center}
\includegraphics[width=6cm]{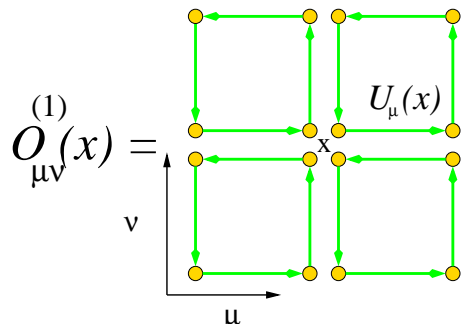}
\caption{Loops required to construct $F_{\mu\nu}$ }
\label{Fmunufig}
\end{center}
\end{figure}

The name ``clover'' is associated with the SW fermion action due to
the lattice discretisation of the field strength tensor, $F_{\mu\nu}$.
An expression for $F_{\mu\nu}$ is obtained by considering the sum of
the four ${1\times{1}}$ link paths surrounding any lattice site in the
$\mu-\nu$ plane as shown in Fig.~\ref{Fmunufig}. Using the expansion
for the elementary link product (see for example
Ref.~\cite{Rothe,Gupta,sbilson} ), we obtain the lattice expression
for $F_{\mu\nu}$
\begin{eqnarray}
g\, a^2\, F_{\mu\nu}(x) & = & \frac{1}{8 i} \biggl [ 
\left ( {\cal{O}}^{(1)}_{\mu\nu}(x)-{\cal{O}}^{(1)\dagger}_{\mu\nu}(x)
\right ) \nonumber \\
&&\qquad - \frac{1}{3} {\rm Tr} \left ( {\cal{O}}^{(1)}_{\mu\nu}(x) -
{\cal{O}}^{(1)\dagger}_{\mu\nu}(x) \right) \biggr ] \, ,
\label{Fmunu}
\end{eqnarray}
where $F_{\mu\nu}$ is made traceless by subtracting $1/3$ of the
trace from each diagonal element
and
\begin{eqnarray}
\lefteqn{ {\cal{O}}^{(1)}_{\mu\nu}(x) = U_{\mu}(x)\, U_{\nu}(x+\hat{\mu})\,
             U^{\dagger}_{\mu}(x+\hat{\nu}) \, U^\dagger_{\nu}(x) }
             \nonumber \\ 
             &+& U_{\nu}(x)\,
             U^{\dagger}_{\mu}(x+\hat{\nu}-\hat{\mu})\, 
             U^{\dagger}_{\nu}(x-\hat{\mu})\, U_{\mu}(x-\hat{\mu})
             \nonumber \\ 
             &+& U^{\dagger}_{\mu}(x-\hat{\mu})\,
             U^{\dagger}_{\nu}(x-\hat{\mu}-\hat{\nu})\,
             U_{\mu}(x-\hat{\mu}-\hat{\nu})\, U_{\nu}(x-\hat{\nu})
             \nonumber \\ 
             &+& U^{\dagger}_{\nu}(x-\hat{\nu})\,
             U_{\mu}(x-\hat{\nu})\, U_{\nu}(x+\hat{\mu}-\hat{\nu})\,
             U^{\dagger}_{\mu}(x) \, .
\label{Omunu}
\end{eqnarray}
Substantial progress has been made to improve $F_{\mu\nu}$ to ${\cal
O}(a^6)$ by adding terms constructed using larger loops
\cite{sbilson}.

\section{Fat-Link Irrelevant Fermion Action}
\label{FLinks}

The established approach to nonperturbative (NP) 
improvement \cite{Luscher:1996sc} tunes the coefficient of the clover
operator to all powers in $g^2$.  Unfortunately, this formulation of
the clover action is susceptible to the problem of exceptional
configurations as the quark mass becomes small.  Chiral symmetry
breaking in the clover fermion action introduces an additive mass
renormalisation into the Dirac operator that can give rise to
singularities in quark propagators at small quark masses.  In
practice, this prevents the simulation of small quark masses and the
use of coarse lattices ($\beta < 5.7 \sim a > 0.18$~fm)
\cite{DeGrand:1999gp,Bardeen:1998gv}.
Furthermore, the plaquette version of $F_{\mu\nu}$, which is commonly
used in Eq.~(\ref{clover}), has large ${\cal O}(a^2)$ errors, which
can lead to errors of the order of $10 \%$ in the topological charge
even on very smooth configurations \cite{Bonnet:2000dc}.

The idea of using fat links in fermion actions was first explored by
the MIT group \cite{Chu:1994vi} and more recently has been studied by
DeGrand {\it et al}.  \cite{DeGrand:1999gp,DeGrand:1998jq,SDDH}, who
showed that the exceptional configuration problem can be overcome by
using a fat-link (FL) clover action.  Moreover, the renormalisation of
the coefficients of action improvement terms is small.
In principle it is acceptable to smear the links of the relevant
operators. The symmetry of the APE smearing process ensures that
effects are ${\cal O}(a^2)$. The factors multiplying the link and
staple ensure the leading order term is $e^{iagA_\mu}$, an element of
SU(3). Issues of projecting the smeared links to SU(3) are ${\cal
O}(a^2)$ effects and therefore correspond to irrelevant operators
\cite{Bernard:1999kc}.  However, the net effect of APE smearing the
links of the relevant operators is to remove gluon interactions at the
scale of the cutoff.  While this has some tremendous benefits, the
short-distance quark interactions are lost. As a result decay
constants and vector-pseudoscalar mass splittings of heavy mesons,
which are sensitive to the wave function at the origin, are suppressed
\cite{Bernard:2002pc}.

The solution to this is to work with two sets of links in the
fermion action.  In the relevant dimension-four operators, one works
with the untouched links generated via Monte Carlo methods, while the
smeared fat links are introduced only in the higher dimension
irrelevant operators.  The effect this has on decay constants and
vector-pseudoscalar mass splittings of heavy mesons is under
investigation and will be reported elsewhere.

Fat links \cite{DeGrand:1999gp,DeGrand:1998jq} are created by
averaging or smearing links on the lattice with their nearest
neighbours in a gauge covariant manner (APE smearing).  The smearing
procedure \cite{Falcioni:1985ei,Albanese:1987ds} replaces a link,
$U_{\mu}(x)$, with a sum of the link and $\alpha$ times its staples
\begin{eqnarray}
U_{\mu}(x)\ \rightarrow\ U_{\mu}'(x) &=&
(1-\alpha)\, U_{\mu}(x)
+ \frac{\alpha}{6} \sum_{\nu=1 \atop \nu\neq\mu}^{4}
  \Big[ U_{\nu}(x)\,
        U_{\mu}(x+\hat{\nu})\,
        U_{\nu}^{\dag}(x+\hat{\mu})\,                       \nonumber \\
      &+& U_{\nu}^{\dag}(x-\hat{\nu})\,
        U_{\mu}(x-\hat{\nu})\,
        U_{\nu}(x-\hat{\nu} +\hat{\mu})
  \Big] \,,
\end{eqnarray} 
followed by projection back to SU(3). We select the unitary matrix
$U_{\mu}^{\rm FL}$ which maximises
$$
{\cal R}e \, {\rm{tr}}(U_{\mu}^{\rm FL}\, U_{\mu}'^{\dagger})\, ,
$$ by iterating over the three diagonal SU(2) subgroups of SU(3).
Performing eight iterations over these subgroups gives gauge
invariance up to seven significant figures.  We repeat the combined
procedure of smearing and projection $n$ times.  We create our fat
links by setting $\alpha = 0.7$ and comparing $n=4$ and 12 smearing
sweeps.
The mean-field improved FLIC action now becomes
\be
S_{\rm SW}^{\rm FL}
= S_{\rm W}^{\rm FL} - \frac{i\, g\, C_{\rm SW}\, \kappa\, r}{2\,
  (u_{0}^{\rm FL})^4}\,
             \bar{\psi}(x)\, \sigma_{\mu\nu}\, F_{\mu\nu}\, \psi(x)\ ,
\label{FLIC}
\ee
where $F_{\mu\nu}$ is constructed using fat links, $u_{0}^{\rm FL}$ is
calculated in an analogous way to Eq.~(\ref{uzero}) with fat links,
and where the mean-field improved Fat-Link Irrelevant Wilson action is
\begin{eqnarray}
S_{\rm W}^{\rm FL}
 =  \sum_x \bar{\psi}(x)\, \psi(x) 
&+& \kappa\, \sum_{x,\mu} \bar{\psi}(x)
    \bigg[ \gamma_{\mu}\,
      \bigg( \frac{U_{\mu}(x)}{u_0}\, \psi(x+\hat{\mu})
        - \frac{U^{\dagger}_{\mu}(x-\hat{\mu})}{u_0}\, \psi(x-\hat{\mu})
      \bigg)                                            \nonumber\\
&-& r\, \bigg(
          \frac{U_{\mu}^{\rm FL}(x)}{u_0^{\rm  FL}}\, \psi(x+\hat{\mu})
        + \frac{U^{{\rm FL}\dagger}_{\mu}(x-\hat{\mu})}{u_0^{\rm FL}}\,
          \psi(x-\hat{\mu})
      \bigg)
    \bigg]\ .
\label{MFIW}
\end{eqnarray}
%

As reported in Table~\ref{meanlink}, the mean-field improvement
parameter for the fat links is very close to 1.  Hence, the mean-field
improved coefficient for $C_{\rm SW}$ is accurate.{\footnote{Our
experience with topological charge operators suggests that it is
advantageous to include $u_0$ factors, even as they approach 1.}}  It
is in this way that the extensive task of non-perturbatively
calculating the renormalisations of the improvement coefficients
discussed in Sec.~\ref{IMPfermions} is avoided.  APE smearing the
links of dimension five operators suppresses the renormalisation,
allowing the precise matching of improvement coefficients with only
tree-level knowledge of their values.

In addition, one can now use highly improved definitions of
$F_{\mu\nu}$ (involving terms up to $u_0^{12})$, which give impressive
near-integer results for the topological charge \cite{sbilson}.
In particular, we employ the 3-loop ${\cal O}(a^4)$-improved definition of
$F_{\mu\nu}$ in which the standard clover-sum of four $1 \times 1$
loops lying in the $\mu ,\nu$ plane is combined with $2 \times 2$ and $3
\times 3$ loop clovers.
Bilson-Thompson {\it et al.} \cite{sbilson} find
\be
g\, F_{\mu\nu} = {-i\over{8}} \left[\left( {3\over{2}}W^{1 \times 1}_{\mu\nu}-
    {3\over{20u_0^4}}W^{2 \times 2}_{\mu\nu}
+{1\over{90u_0^8}}W^{3 \times 3}_{\mu\nu}\right) - {\rm
h.c.}\right]_{\rm Traceless}\ 
\ee
where $W^{n \times n}$ is the clover-sum of four $n \times n$ loops
and $F_{\mu\nu}$ is made traceless by subtracting $1/3$ of the trace from
each diagonal element of the $3 \times 3$ colour matrix.
This definition reproduces the continuum limit with ${\cal O}(a^6)$
errors.
On approximately self-dual configurations, this operator produces integer 
topological charge to better than 4 parts in $10^4$. We also consider
a 5-loop improved $F_{\mu\nu}$ for the $20^3 \times 40$ lattice at
$\beta=4.53$. Since the results for the 5-loop operator agree with the
3-loop version to better than 4 parts in $10^4$ \cite{sbilson}, we are
effectively using the same action as far the scaling analysis is concerned.

\begin{table}[t]
\begin{center}
\caption{The value of the mean link for different numbers of APE smearing
  sweeps, $n$, at $\alpha = 0.7$ on a $16^3 \times 32$ lattice at
  $\beta=4.60$ which corresponds to a lattice spacing of 0.122(2)~fm set
  by the string tension. \label{meanlink}}
\vspace*{0.5cm}
\begin{tabular}{cc|ccc|cc}
$n$ & & & $u^{\rm FL}_0$ & & & $(u^{\rm FL}_0)^4$ \\ \hline
    0 & & & 0.88894473 & & & 0.62445197 \\
    4 & & & 0.99658530 & & & 0.98641100 \\
    12& & & 0.99927343 & & & 0.99709689
\end{tabular}
\end{center}
\end{table}

The use of thin links in Eq. (\ref{MFIW}) ensures that the relevant
dimension-four operators see all the dynamics of the Monte-Carlo
generated gauge fields.  Upon expanding the thin links in terms of the
gauge potential, ${\mathcal O}(a^2)$ contributions of energy-dimension
six are revealed, which, ideally, should be removed via the fat-link
irrelevant operator procedure.  
Fortunately, actions with many irrelevant operators ({\it e.g.} the
D$_{234}$ action) can now be handled with confidence as tree-level
knowledge of the improvement coefficients is sufficient.
However, as we will see, the scaling of the ${\mathcal O}(a)$-improved
FLIC fermion action of Eqs. (\ref{FLIC}) and (\ref{MFIW}) is excellent
already, and the added computational expense of the two-link hopping
terms required in ${\mathcal O}(a^2)$-improvement is not well
motivated at present.


Work by DeForcrand {\it et al}. \cite{deForcrand:1997sq} suggests that 7
cooling sweeps are required to approach topological charge within 1$\%$ of
integer value.
This is approximately 16 APE smearing sweeps at $\alpha = 0.7$
\cite{Bonnet:2001rc}.
However, achieving integer topological charge is not necessary for the
purposes of studying hadron masses, as has been well established.
To reach integer topological charge, even with improved definitions of
the topological charge operator, requires significant smoothing and
associated loss of short-distance information.
Instead, we regard this as an upper limit on the number of smearing
sweeps.

Using unimproved gauge fields and an unimproved topological charge
operator, Bonnet {\it et al}. \cite{Bonnet:2000dc} found that the
topological charge settles down after about 10 sweeps of APE smearing at
$\alpha=0.7$.
Consequently, we create fat links with APE smearing parameters $n=12$ and
$\alpha=0.7$.
This corresponds to $\sim 2.5$ times the smearing used in
Refs.~\cite{DeGrand:1999gp,DeGrand:1998jq}.
Further investigation reveals that improved gauge fields with a small
lattice spacing ($a=0.122$~fm) are smooth after only 4 sweeps.
Hence, we perform calculations with 4 sweeps of smearing at $\alpha=0.7$
and consider $n=12$ as a second reference. 
Table~\ref{meanlink} lists the values of $u_0^{\rm FL}$ for $n=0$, 4 and
12 smearing sweeps.

We also compare our results with the standard Mean-Field Improved
Clover (MFIC) action. We mean-field improve as defined in
Eqs.~(\ref{FLIC}) and (\ref{MFIW}) but with thin links throughout. For
this action, the standard 1-loop definition of $F_{\mu\nu}$ is
used.

\section{Lattice Simulations}
\label{simulations}

The simulations are performed using the Luscher-Weisz
\cite{Luscher:1984xn} mean-field improved, plaquette plus rectangle,
gauge action on $12^3 \times 24$ and $16^3 \times 32$ lattices with
lattice spacings of 0.093, 0.122 and 0.165~fm determined from the
string tension with $\sqrt\sigma=440$~MeV. We define
\be
S_{\rm G} = \frac{5\beta}{3}
      \sum_{\rm{sq}}\frac{1}{3}{\cal R}e\ {\rm{tr}}(1-U_{\rm{sq}}(x))
    - \frac{\beta}{12u_{0}^2}
      \sum_{\rm{rect}}\frac{1}{3}{\cal R}e\ {\rm{tr}}(1-U_{\rm{rect}}(x))\ ,
\label{gaugeaction}
\ee
where the operators $U_{\rm{sq}}(x)$ and $U_{\rm{rect}}(x)$ are
defined as
%
\begin{eqnarray}
U_{\rm{sq}}(x)
&=& U_{\mu}(x)\, U_{\nu}(x+\hat{\mu})\,
    U^{\dagger}_{\mu}(x+\hat{\nu})\, U^\dagger_{\nu}(x)\ ,      \\
U_{\rm{rect}}(x)
&=& U_{\mu}(x)\, U_{\nu}(x+\hat{\mu})\,
    U_{\nu}(x+\hat{\nu}+\hat{\mu})\,                    \nonumber\\
& &\times U^{\dagger}_{\mu}(x+2\hat{\nu})\,
         U^{\dagger}_{\nu}(x+\hat{\nu})\,
         U^\dagger_{\nu}(x)\,                           \nonumber\\
&+& U_{\mu}(x)\, U_{\mu}(x+\hat{\mu})\,
    U_{\nu}(x+2\hat{\mu})\,                             \nonumber\\
& &\times U^{\dagger}_{\mu}(x+\hat{\mu}+\hat{\nu})\,
         U^{\dagger}_{\mu}(x+\hat{\nu})\, U^\dagger_{\nu}(x)\ .
\label{actioneqn}
\end{eqnarray}
%
The link product $U_{\rm{rect}}(x)$ denotes the rectangular $1\times2$
and $2\times1$ plaquettes, and for the tadpole improvement factor we
employ the plaquette measure
\be
u_0 = \left( \frac{1}{3}{\cal R}e \, {\rm{tr}}\langle U_{\rm{sq}}\rangle
      \right)^{1/4}\ .
\label{uzero}
\ee

Initial studies of FLIC, mean-field improved clover and Wilson quark
actions were made using 50 configurations. The scaling analysis of
FLIC fermions was performed with
a total of 200 configurations at each lattice spacing and volume.  
In addition, for the light quark simulations, 94 configurations are
used on a $20^3 \times 40$ lattice with $a=0.134(2)$~fm.
Gauge configurations are generated using the Cabibbo-Marinari
pseudo-heat-bath algorithm with three diagonal SU(2) subgroups looped
over twice.
Simulations are performed using a parallel algorithm with appropriate
link partitioning \cite{Bonnet:2000db},
and the error analysis is performed by a
third-order, single-elimination jackknife, with the $\chi^2$ per
degree of freedom ($N_{\rm DF}$) obtained via covariance matrix fits.

A fixed boundary condition is used for the fermions by setting
\be
U_t (\vec{x},nt) = 0 \qquad {\rm and} \qquad U_t^{\rm FL} (\vec{x},nt) = 0 
\qquad \forall\ \vec{x}\ 
\ee
in the hopping terms of the fermion action.  The fermion source is
centered at the space-time location {$(x,y,z,t) = (1,1,1,3)$}, which
allows for two steps backward in time without loss of signal, for all
simulations except those on the $20^3\times 40$ lattice at
$\beta=4.53$ which has the fermion source located at $(x,y,z,t) =
(1,1,1,8)$.
Gauge-invariant Gaussian smearing \cite{Gusken:1990qx} in the spatial
dimensions is applied at the source to increase the overlap of the
interpolating operators with the ground states.
The source-smearing technique \cite{Gusken:1990qx} starts with a point
source, 
\be
\psi_{0\, \alpha}^{\phantom{0}\, a}({\vec x}, t) = \delta^{ac}
\delta_{\alpha\gamma} \delta_{{\vec x},{\vec x}_0} \delta_{t,t_0}
\label{ptsource}
\ee
for source colour $c$, Dirac $\gamma$, position ${\vec x}_0 = (1,1,1)$
and time $t_0$ and proceeds via the iterative scheme,
\[
\psi_i({\vec x},t) = \sum_{{\vec x}'} F({\vec x},{\vec x}') \, \psi_{i-1}({\vec x}',t) \, ,
\]
where
\[
F({\vec x},{\vec x}') = \frac{1}{(1+\alpha)} \left( \delta_{{\vec x},
    {\vec x}'} + 
  \frac{\alpha}{6} \sum_{\mu=1}^3 \left [ U_\mu({\vec x},t) \, \delta_{{\vec x}',
{\vec x}+\widehat\mu} + 
U_\mu^\dagger({\vec x}-\widehat\mu,t) \, \delta_{{\vec x}', {\vec
    x}-\widehat\mu} \right ] \right) \, . 
\]
Repeating the procedure $N$ times gives the resulting fermion source
\be
\psi_N({\vec x},t) = \sum_{{\vec x}'} F^N({\vec x},{\vec x}') \, \psi_0({\vec x}',t) \, .
\ee
The parameters $N$ and $\alpha$ govern the size and shape of the
smearing function. We simulate with $N=20$ and $\alpha=6$, except on
the $20^3\times40$ lattice which has $N=35$.
The propagator, $S$, is obtained from the smeared source by
solving 
\be
M_{\alpha\beta}^{ab}\, S_{\beta\gamma}^{bc} =
\psi_{\alpha}^{a}\, ,
\ee
for each colour, Dirac source $c,\, \gamma$ respectively of
Eq.~(\ref{ptsource}) via the BiStabilised Conjugate Gradient algorithm
\cite{BiCG}.

\section{Scaling of FLIC Fermions}
\label{discussion}

Hadron masses are extracted from the Euclidean time dependence of the
calculated two-point correlation functions.
For baryons the correlation functions are given by
\be
 G(t;\vec{p},\Gamma)
= \sum_x e^{-i\, \vec p\cdot \vec x}\ \Gamma^{\beta\alpha}
\langle\Omega|T[\chi^{\alpha}(x)\bar{\chi}^{\beta}(0)]|\Omega\rangle\ ,
\ee
where $\chi$ are standard baryon interpolating fields, $\Omega$
represents the QCD vacuum, $\Gamma$ is a $4\times4$ matrix in Dirac
space, and $\alpha$, $\beta$ are Dirac indices.
At large Euclidean times one has
\be
G(t;\vec{p},\Gamma)
\simeq \frac{Z^2}{2 E_p}\ e^{-E_p t}\
{\rm tr}\left[ \Gamma(-i\gamma\cdot p+M)\right]\ ,
\ee
where $Z$ represents the coupling strength of $\chi(0)$ to the baryon,
and $E_p = ({\vec p}^{\, 2} + M^2 )^{1/2}$ is the energy.
Selecting ${\vec p}=0$ and $\Gamma = (1+\gamma_4)/4$, the effective
baryon mass is then given by
\be
M(t) = \log [G(t)] - \log[G(t+1)]\ .
\ee
Meson masses are determined via analogous standard procedures.  The
critical value of $\kappa$, $\kappa_{\rm cr}$, is determined by
linearly extrapolating $m_{\pi}^2$ as a function of $m_q$ to zero.
%

\begin{figure}[t]
\begin{center}
\includegraphics[height=0.95\hsize,angle=90]{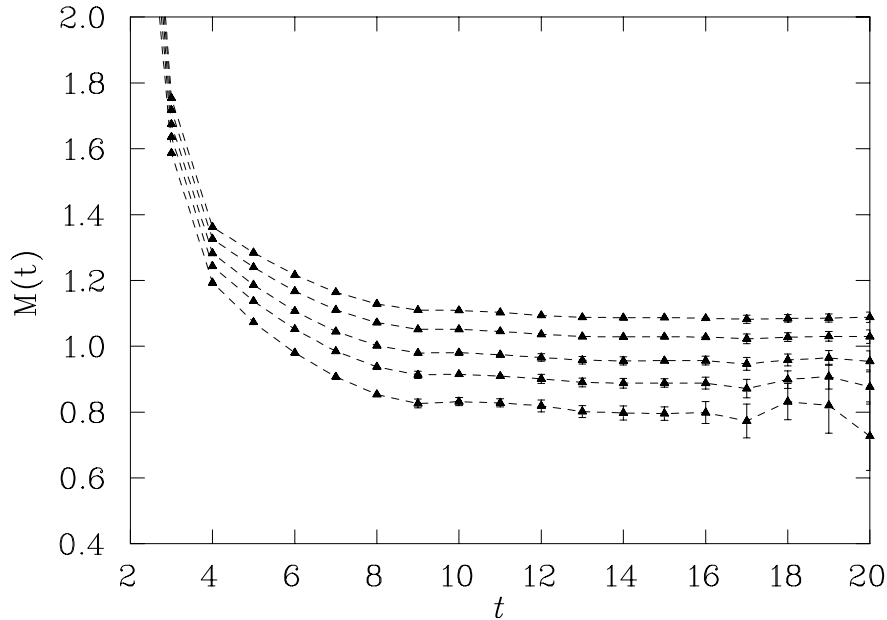} 
\vspace*{0.5cm}
\caption{Effective mass plot for the nucleon for the FLIC action from
200 configurations on a $16^3 \times 32$ lattice at $\beta=4.60$ which
corresponds to a lattice spacing of 0.122(2)~fm set by the string
tension.  The fat links are created with 4 sweeps of smearing at
$\alpha = 0.7$.  The five sets of points correspond to the $\kappa$
values listed in Table~\protect\ref{masses}, with $\kappa$ increasing
from top down.}
\label{rho}
\end{center}
\end{figure}

Figure~\ref{rho} shows the nucleon effective mass plot for the FLIC
action on a $16^3 \times 32$ lattice at $\beta=4.60$ which corresponds to
a lattice spacing of 0.122(2)~fm set by the string tension. The fat links
are created with 4 APE smearing sweeps at $\alpha=0.7$ (``FLIC4'').
The effective mass plots for the other hadrons are similar, and all
display acceptable plateau behavior.  Good values of $\chi^2 / N_{\rm
  DF}$ are obtained for many different time-fitting intervals as long
as one fits after time slice 8.  All fits for this action are
therefore performed on time slices 9 through 14.  For the Wilson
action and the FLIC action with $n=12$ (``FLIC12''), the effective
mass plots look similar to Fig.~\ref{rho} and display good plateau
behavior. The fitting
regimes used for these actions are 9-13 and 9-14, respectively.

\begin{table}[t]
\begin{center}
\caption{Values of $\kappa$ and the corresponding $\pi,\, \rho,\,
  {\rm N}$ and $\Delta$ masses for the FLIC action with 4 sweeps of
  smearing at $\alpha=0.7$ on a $16^3 \times 32$ lattice at
  $\beta=4.60$. The value for $\kappa_{\rm cr}$ is provided
  in Table~\protect\ref{kappaT}.  A string tension analysis
  incorporating the lattice coulomb term provides $a=0.122(2)$ fm for
  $\sqrt\sigma=440$~MeV.\label{masses}}
\vspace*{0.5cm}
  \begin{tabular}{cc|ccc|ccc|ccc|cc}
        $\kappa$ & & &
        $m_{\pi}\, a$ & & & $m_{\rho}\, a$ & & &
        $m_{\rm N}\, a$ & & & $m_{\Delta}\, a$      \\ \hline
    0.1260 & & & 0.5797(23) & & & 0.7278(39) & & & 1.0995(58)  & & & 1.1869(104) \\
    0.1266 & & & 0.5331(24) & & & 0.6951(45) & & & 1.0419(64)  & & & 1.1387(121) \\
    0.1273 & & & 0.4744(27) & & & 0.6565(54) & & & 0.9709(72)  & & & 1.0816(152) \\
    0.1279 & & & 0.4185(30) & & & 0.6229(65) & & & 0.9055(82)  & & & 1.0310(194) \\
    0.1286 & & & 0.3429(37) & & & 0.5843(97) & & & 0.8220(102) & & & 0.9703(286) \\
\end{tabular}
\end{center}
\end{table}

\begin{table}[t]
\begin{center}
\caption{Values of $\kappa$ and $\kappa_{\rm cr}$ for the four
  different actions on a $16^3 \times 32$ lattice at $\beta=4.60$ which
  corresponds to a lattice spacing of 0.122(2)~fm set by the string
  tension.\label{kappaT}}
\vspace*{0.5cm}
  \begin{tabular}{cc|ccc|ccc|ccc|cc}
        & & & Wilson & & & FLIC12 & & & FLIC4 & & & MFIC        \\ \hline
    $\kappa_1$ & & & 0.1346 & & & 0.1286 & & & 0.1260 & & & 0.1196 \\
    $\kappa_2$ & & & 0.1353 & & & 0.1292 & & & 0.1266 & & & 0.1201 \\
    $\kappa_3$ & & & 0.1360 & & & 0.1299 & & & 0.1273 & & & 0.1206 \\
    $\kappa_4$ & & & 0.1367 & & & 0.1305 & & & 0.1279 & & & 0.1211 \\
    $\kappa_5$ & & & 0.1374 & & & 0.1312 & & & 0.1286 & & & 0.1216 \\ \hline
    $\kappa_{\rm cr}$ & & & 0.1390 & & & 0.1328 & & & 0.1300 & & & 0.1226

\end{tabular}

\end{center}
\end{table}

The values of $\kappa$ used in the simulations for all quark actions
are given in Table~\ref{kappaT}. We have also provided the values of
$\kappa_{\rm cr}$ for these fermion actions when using our mean-field 
improved, plaquette plus rectangle, gauge action at $\beta = 4.60$. 
We have mean-field improved our fermion actions so we expect
the values for $\kappa_{\rm cr}$ to be close to the tree-level value of 0.125.
Improved chiral properties are seen for the FLIC and MFIC actions, with FLIC4
performing better than FLIC12.

\begin{figure}[t]
\begin{center}
\includegraphics[height=0.95\hsize,angle=90]{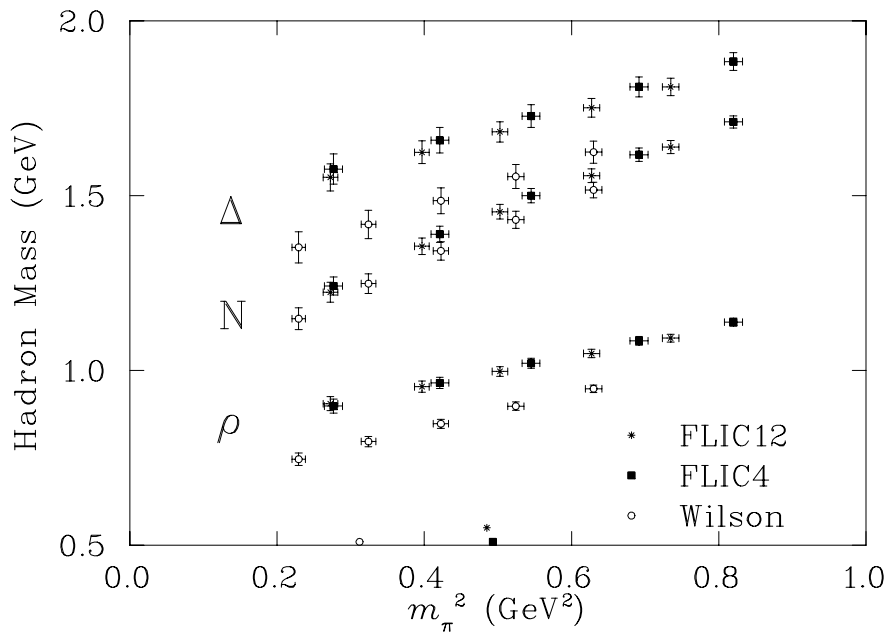} 
\vspace*{0.5cm}
\caption{Masses of the nucleon, $\Delta$ and $\rho$ meson versus
  $m_{\pi}^2$ for the FLIC4, FLIC12 and Wilson actions on a $16^3 \times
  32$ lattice at $\beta=4.60$ which corresponds to a lattice spacing
  of 0.122(2)~fm set by the string tension.}
\label{Mvsmpi2}
\end{center}
\end{figure}

The behavior of the $\rho$, nucleon and $\Delta$ masses as a function of
squared pion mass are shown in Fig.~\ref{Mvsmpi2} for the various actions.
The first feature to note is the excellent agreement between the FLIC4
and FLIC12 actions.
On the other hand, the Wilson action appears to lie somewhat low in
comparison.
It is also reassuring that all actions give the correct mass ordering in
the spectrum.
The value of the squared pion mass at $m_{\pi}/m_{\rho} = 0.7$ is plotted
on the abscissa for the three actions as a reference point.
This point is chosen in order to allow comparison of different results by
interpolating them to a common value of $m_{\pi}/m_{\rho} = 0.7$, rather
than extrapolating them to smaller quark masses, which is subject to
larger systematic and statistical uncertainties.

\begin{figure}[t]
\begin{center}
\includegraphics[angle=90,height=11cm]{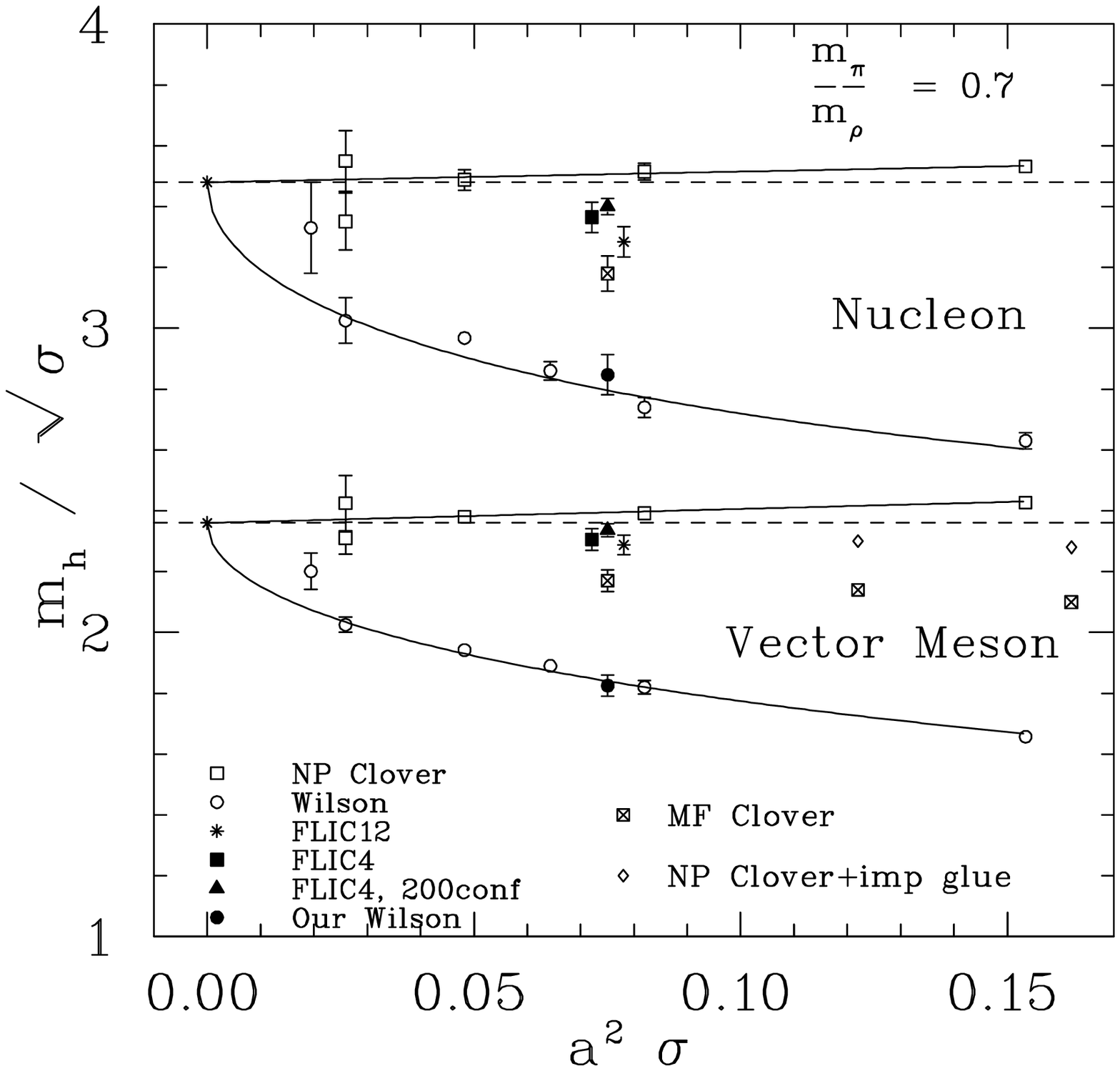} 
\vspace*{0.5cm}
\caption{Nucleon and vector meson masses for the Wilson, NP-improved,
mean-field clover and FLIC actions.  Results from the present
simulations, based on 50 configurations and indicated by the solid
points, are obtained by interpolating the results of
Fig.~{\protect\ref{Mvsmpi2}} to $m_\pi / m_\rho = 0.7$.  The fat links
are constructed with $n=4$ (solid squares) and $n=12$ (stars) smearing
sweeps at $\alpha = 0.7$.  The solid triangles are results for the
FLIC4 action when 200 configurations are used in the analysis.  The
FLIC results are offset from the central value for clarity.  Our MF
clover result at $a^2 \sigma \sim 0.075$ lies systematically low
relative to the FLIC actions.}
\label{scaling1}
\end{center}
\end{figure}

The scaling behaviour of the different actions is illustrated in
Fig.~\ref{scaling1}.
The present results for the Wilson action agree with those of
Ref.~\cite{Edwards:1998nh}.
The first feature to observe in Fig.~\ref{scaling1} is that actions with
fat-link irrelevant operators perform extremely well.
For both the vector meson and the nucleon, the FLIC actions perform
significantly better than the mean-field improved clover action.
It is also clear that the FLIC4 action performs systematically better
than the FLIC12.
This suggests that 12 smearing sweeps removes too much short-distance
information from the gauge-field configurations.
On the other hand, 4 sweeps of smearing combined with our
${\cal O} (a^4)$ improved $F_{\mu\nu}$ provides excellent results,
without the fine tuning of $C_{\rm SW}$ in the NP improvement program.

Notice that for the $\rho$ meson, a linear extrapolation of previous
mean-field improved clover results in Fig.~\ref{scaling1} passes through
our mean-field improved clover result at $a^2 \sigma \sim 0.075$ which 
lies systematically low relative to the FLIC actions.
However, a linear extrapolation does not pass through the continuum limit
result, thus confirming the presence of significant ${\cal O}(a)$ errors
in the mean-field improved clover fermion action.
%
%
While there are no NP-improved clover plus improved glue simulation
results at $a^2 \sigma \sim 0.075$, the simulation results that are
available indicate that the fat-link results also compete well with
those obtained with a NP-improved clover fermion action.

Having determined FLIC4 is the preferred action, we have increased the
number of configurations to 200 for this action.  As expected, the
error bars are halved and the central values for the FLIC4 points move
to the upper end of the error bars on the 50 configuration result,
further supporting the promise of excellent scaling.

\begin{table}[t]
\begin{center}
\caption{String tensions, $\beta$, volumes and results for the vector
  meson and nucleon masses interpolated to $m_P / m_V = 0.7$. The scale for
  the small $\beta=4.60$ lattice estimates are taken from the large
  $\beta=4.60$ lattice \label{lattices}}
  \begin{tabular}{cc|ccc|ccc|ccc|ccc|ccc|cc}
        $\beta$ & & & Volume & & & $N_{\rm configs}$ & & & $a\sqrt{\sigma}$ & & & 
        $m_v / \sqrt{\sigma}$ & & & $m_N / \sqrt{\sigma}$ & & & $u_0$     \\ \hline
    4.38  & & & $16^3\times 32$ & & & 200 & & & 0.371 & & & 2.378(25) 
& & & 3.450(35) & & & 0.8761 \\
    4.53  & & & $20^3\times 40$ & & & 94 & & & 0.299 & & & 2.318(18) 
& & & 3.408(26) & & & 0.8859 \\
    4.60  & & & $12^3\times 24$ & & & 200 & & & 0.274 & & & 2.434(26) 
& & & 3.554(33) & & & 0.8889 \\
    4.60  & & & $16^3\times 32$ & & & 200 & & & 0.274 & & & 2.336(22) 
& & & 3.400(26) & & & 0.8889 \\
    4.80  & & & $16^3\times 32$ & & & 200 & & & 0.210 & & & 2.427(23) 
& & & 3.538(61) & & & 0.8966 \\
\end{tabular}
\end{center}
\end{table}

\begin{figure}[t]
\begin{center}
{\includegraphics[angle=90,height=11cm]{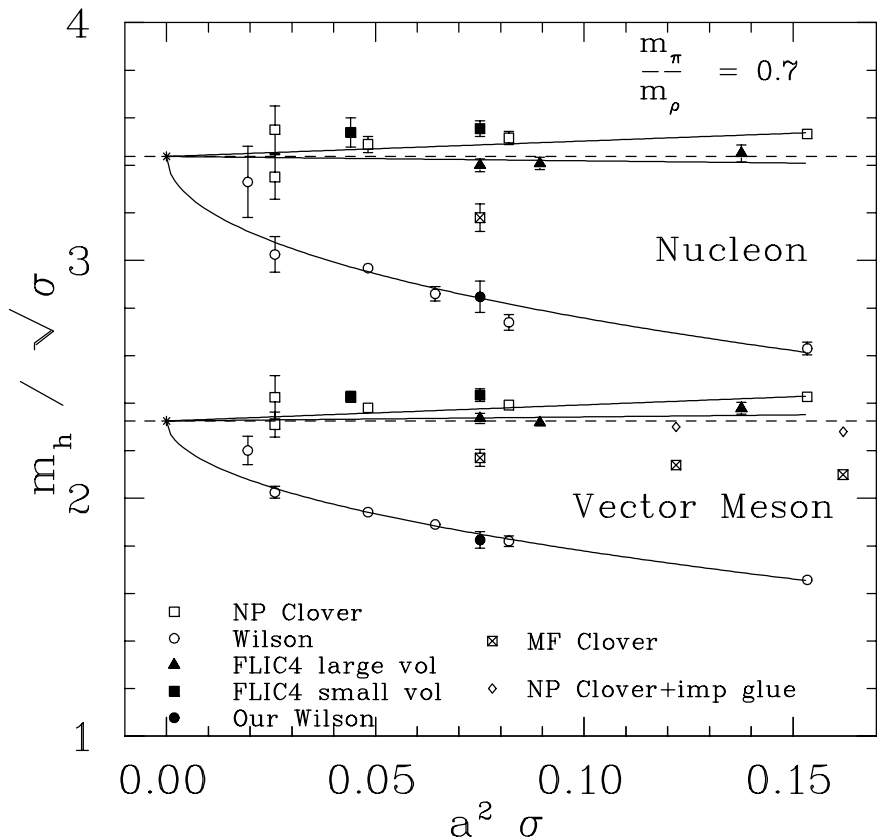}}
\vspace*{0.5cm}
\caption{Nucleon and vector meson masses for the Wilson, Mean-Field
  (MF) improved clover, NP-improved clover and FLIC actions obtained by
  interpolating simulation results to $m_\pi / m_\rho = 0.7$.  For the
  FLIC action (``FLIC4''), fat links are constructed with $n=4$
  APE-smearing sweeps with smearing fraction $\alpha = 0.7$, except
  for the point at $a^2 \sigma \sim 0.09$ which has $n=6$.  Results
  from the current simulations are indicated by the solid
  symbols; those from earlier simulations by open or hatched symbols.
  The solid-lines illustrate fits, constrained to have a common
  continuum limit, to FLIC, NP-improved clover and Wilson fermion
  action results obtained on physically large lattice volumes. }
\label{scalefit} 
\end{center}
\end{figure}

In order to further test the scaling of the FLIC action at different
lattice spacings, we consider four different lattice spacings and
three different volumes. String tensions, volumes and hadron masses
are given in Table~\ref{lattices} and the results are displayed in
Fig.~\ref{scalefit}.
The two different volumes used at $a^2 \sigma \sim 0.075$ indicate a
small finite volume effect, which increases the mass for the smaller
volume at $a^2 \sigma \sim 0.075$ and $\sim 0.045$.
Examination of points from the small and large volumes separately
indicates continued scaling toward the continuum limit. While the
finite volume effect will produce a different continuum limit value,
the slope of the points from the smaller and larger volumes agree,
consistent with errors of ${\cal O}(a^2)$.  

Focusing on simulation results from physical volumes with extents
$\sim 2$ fm and larger, we perform a simultaneous fit of the FLIC,
NP-improved clover and Wilson fermion action results.  The fits are
constrained to have a common continuum limit and assume errors are
${\cal O} (a^2)$ for FLIC and NP-improved clover actions and ${\cal O}
(a)$ for the Wilson action.  An acceptable $\chi^2$ per degree of
freedom is obtained for both the nucleon and $\rho$-meson fits.  These
results indicate that FLIC fermions provide a new form of
nonperturbative ${\cal O}(a)$ improvement.  The FLIC fermion results
display nearly perfect scaling indicating ${\cal O} (a^2)$ errors are
small for this action.

\begin{figure}[t]
\begin{center}
\includegraphics[height=0.90\hsize,angle=90]{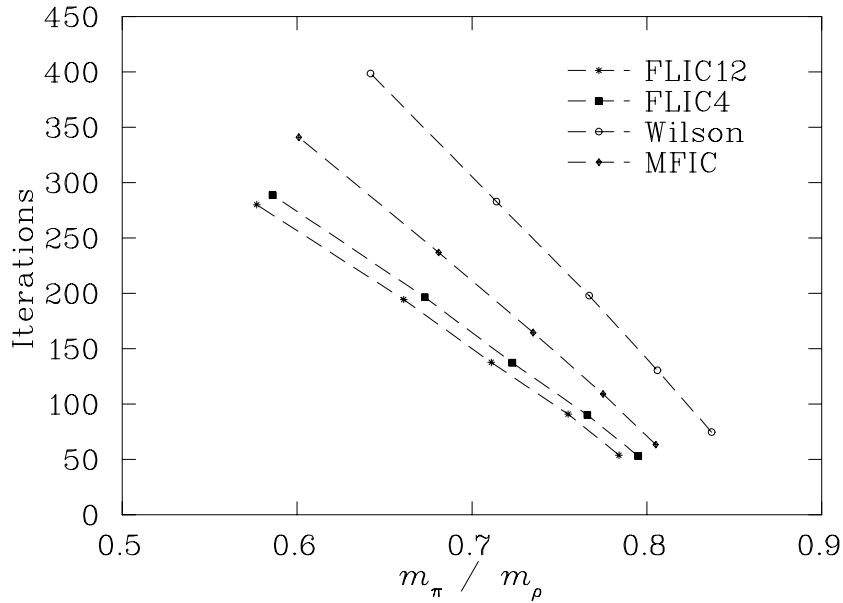} 
\vspace*{0.5cm} 
\caption{Average number of stabilised biconjugate gradient iterations for
  the Wilson, FLIC and mean-field improved clover (MFIC) actions
  plotted against $m_{\pi} / m_{\rho}$.  The fat links are constructed
  with $n=4$ (solid squares) and $n=12$ (stars) smearing sweeps at
  $\alpha = 0.7$ on a $16^3 \times 32$ lattice at $\beta=4.60$ which
  corresponds to a lattice spacing of 0.122(2)~fm set by the string
  tension.}
\label{iterations}
\end{center}
\end{figure}

\section{Search For Exceptional Configurations}
\label{exceptionalSection}

Chiral symmetry breaking in the Wilson action allows continuum zero
modes of the Dirac operator to be shifted into the negative mass
region.
This problem is accentuated as the gauge fields become rough ($a \to$
large).  Local lattice artifacts at the scale of the cutoff (often
referred to as dislocations) give rise to spurious near zero modes.
The quark propagator can then encounter singular behaviour as the
quark mass becomes light.

Exceptional configurations are a severe problem in quenched QCD (QQCD)
because instantons are low action field configurations which appear
readily in QQCD. These instanton configurations give rise to
approximate zero modes which should be suppressed at light quark
masses by ${\rm det}\, M$ which is present in the link
updates in full QCD. 
This determinant is not present in QQCD and as a result,
near-zero modes are overestimated in the ensemble.

The addition of the clover term to the fermion action broadens the
distribution of near-zero modes.
As a result, the clover
action is notorious for revealing the exceptional configuration
problem in QQCD. The FLIC action is expected to reduce the number of
exceptional configurations by smoothing the gauge fields of the
irrelevant operator via APE
smearing \cite{Falcioni:1985ei,Albanese:1987ds}. The smoothing
procedure has the effect of suppressing the local lattice artifacts and
narrowing the distribution of near-zero modes, enabling
simulations to be performed at light quark masses not currently
accessible with the standard mean-field or non-perturbative improved
clover fermion actions.

\begin{figure}[t]
\begin{center}
\includegraphics[height=0.90\hsize,angle=90]{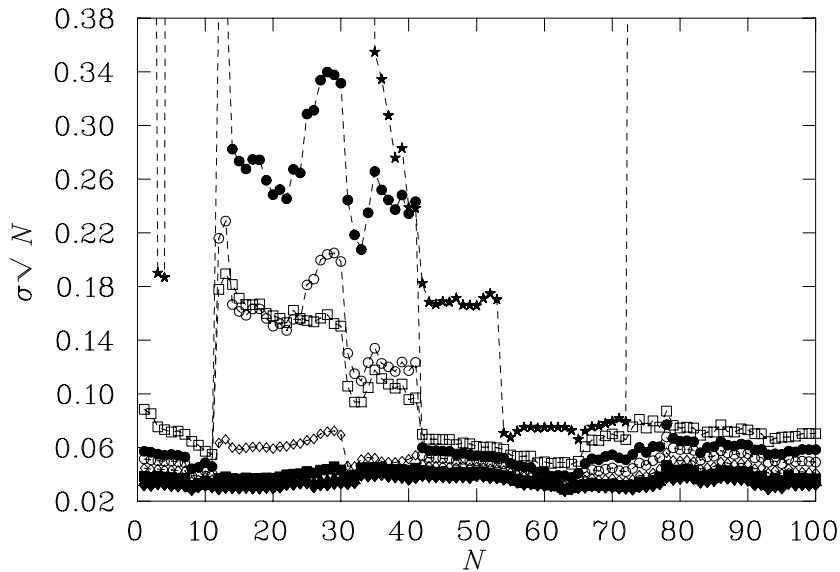}
\vspace*{0.5cm}
\caption{ The standard deviation in the error of the $\pi$ mass
  for eight quark masses (with the star symbols being the lightest
  quark mass) calculated on 30 configurations plotted against the
  starting configuration number for the FLIC-fermion action on a $20^3
  \times 40$ lattice with $a=0.134(2)$~fm.  }
\label{PiErrvsCFG}
\end{center}
\end{figure}

In order to access the light quark regime, we would like our preferred
action to be efficient when inverting the fermion matrix.
Fig.~\ref{iterations} compares the convergence rates of the different
actions on a $16^3 \times 32$ lattice at $\beta=4.60$ by plotting the
number of stabilised biconjugate gradient \cite{BiCG} iterations
required to invert the fermion matrix as a function of $m_{\pi} /
m_{\rho}$.  For any particular value of $m_{\pi} / m_{\rho}$, the FLIC
actions converge faster than both the Wilson and mean-field improved
clover fermion actions.  Also, the slopes of the FLIC lines are
smaller in magnitude than those for Wilson and mean-field improved
clover actions, which provides great promise for performing cost
effective simulations at quark masses closer to the physical values.
Problems with exceptional configurations have prevented such
simulations in the past.

The ease with which one can invert the fermion matrix using FLIC
fermions (also see Ref~\cite{WASEEM}) leads us to attempt simulations
down to light quark masses corresponding to $m_{\pi} / m_{\rho} = $
0.35. Previous attempts with Wilson-style fermion actions on
configurations with lattice spacing $\ge 0.1$~fm have only
succeeded in getting down to $m_{\pi} / m_{\rho} = 0.47$ \cite{TMass}.
In order to search for exceptional configurations, we follow the
technique used by Della Morte {\it et al.} \cite{TMass} and note that
in the absence of exceptional configurations, the standard deviation
of an observable will be independent of the number of configurations
considered in the average.  Exceptional configurations reveal
themselves by introducing a significant jump in the standard deviation
as the configuration is introduced into the average.  In severe cases,
exceptional configurations can lead to divergences in correlation
functions or prevent the matrix inversion process from converging.

\begin{figure}[t]
\begin{center}
{\includegraphics[height=0.90\hsize,angle=90]{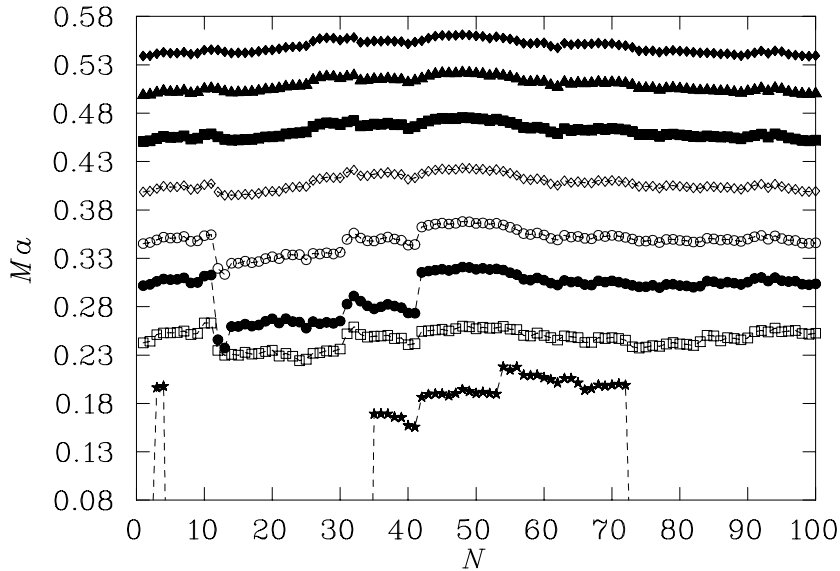}}
\vspace*{0.5cm}
\caption{ The $\pi$ mass
  calculated for eight quark masses (with the star symbols being the
  lightest quark mass) on 30 configurations plotted against the
  starting configuration number for the FLIC-fermion action on a $20^3
  \times 40$ lattice with $a=0.134$~fm.  }
\label{PiMassvsCFG}
\end{center}
\end{figure}

The simulations are on a $20^3 \times 40$ lattice with a lattice
spacing of 0.134(2)~fm set by a string tension analysis incorporating
the lattice coulomb term.  The physical length of the lattice is
$\sim$ 2.7~fm.  We have used an initial set of 100 configurations,
using $n=6$ sweeps of APE-smearing and a five-loop improved lattice
field-strength tensor.  Fig.~\ref{PiErrvsCFG} shows the standard
deviation of the pion mass for eight quark masses on subsets of 30
(consecutive) configurations with a cyclic property enforced from
configuration 100 to configuration 1.  At first glance, it is obvious
that the error blows up for several quark masses at $N=12$ and drops
again at $N=42$. As configurations 12 through 41 are included in the
average at $N=12$, this indicates that configuration number 41 is a
candidate for an exceptional configuration. An inspection of the pion
mass in Fig.~\ref{PiMassvsCFG} shows that the pion mass for the third
lightest quark mass decreases significantly more than the second or
fourth lightest quark masses. This indicates that $\kappa_{\rm cr}$
for this configuration lies somewhere between $\kappa_6$ and
$\kappa_7$. A solution to this problem would be to use the the
modified quenched approximation (MQM) from Ref.~\cite{Bardeen:1998gv}
and move $\kappa_{\rm cr}$ on this configuration back to the ensemble
average for $\kappa_{\rm cr}$.  However, since the movement of
$\kappa_{\rm cr}$ is largely a quenched artifact and would be
suppressed in a full QCD simulation we prefer to simply identify and
remove such configurations from the ensemble.
Obviously, if we find that a significant percentage of our
configurations are having trouble at a particular quark mass, then it
would make no sense to proceed with the simulation. We would then have
to conclude that we have reached the light quark mass limit of our
action and simply step back to the next lightest mass.

\begin{figure}[t]
\begin{center}
{\includegraphics[height=0.90\hsize,angle=90]{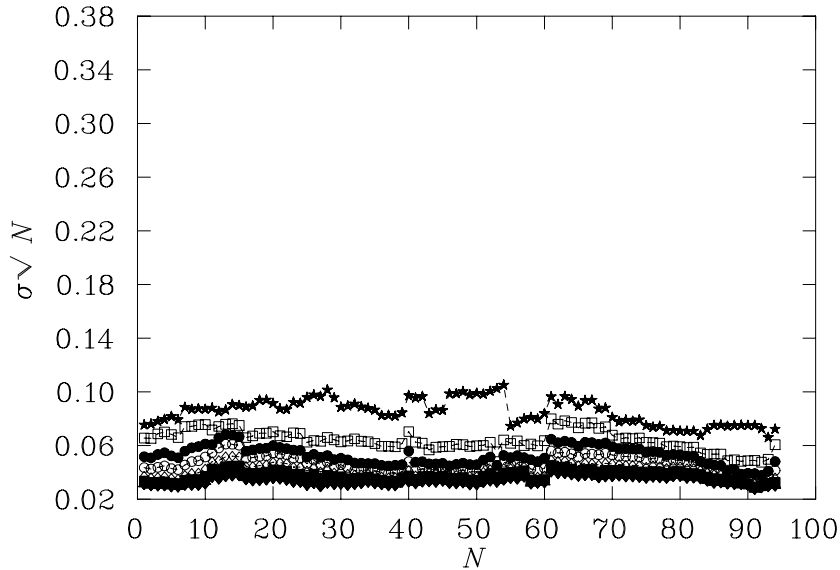}}
\vspace*{0.5cm}
\caption{ The standard deviation in the error of the $\pi$ mass
  for eight quark masses (with the star symbols being the lightest
  quark mass) calculated on 30 configurations plotted against the
  starting configuration number for the FLIC-fermion action on a $20^3
  \times 40$ lattice with $a=0.134$~fm. Configuration numbers 2, 13,
  30, 34, 41 and 53 have been omitted.}
\label{PiErrvsCFG2}
\end{center}
\end{figure}

Now let us return to Fig.~\ref{PiErrvsCFG}. In addition to the highly
exceptional configuration number 41, we also notice a large increase
in error in the lightest quark mass for configuration numbers 2,
13, 30, 34 and 53. Upon removal of these configurations, we see in
Fig.~\ref{PiErrvsCFG2} a near-constant behaviour of the standard
deviation for the remaining configurations. This means
that our elimination rate for our FLIC6 action on a lattice with a
spacing 0.134 fm is about $6\%$. So for the 100 configurations used in
this analysis, we are able to use 94 of them to extract hadron masses.

A similar analysis on a $16^3 \times 32$ lattice at $\beta = 4.60$
providing a finer lattice spacing of 0.122(2) fm reveals a much
smaller exceptional configuration rate.  In a sample of 200
configurations, 4 were identified as exceptional.  The increase from
2\% to 6\% in going from $a \simeq 0.125$ to 0.135 fm suggests that
the coarser lattice spacing is near the limit of applicability for
FLIC fermions in the light quark mass regime.

\begin{table}[t]
\begin{center}
\caption{Values of $\kappa$ and the corresponding $\pi,\, \rho,\, {\rm
        N}$ and $\Delta$ masses on a $20^3 \times 40$ lattice for the
        FLIC action with 6 sweeps of smearing at $\alpha=0.7$.  A
        string tension analysis incorporating the lattice coulomb term
        provides $a=0.134(2)$ fm for
        $\sqrt\sigma=440$~MeV.\label{453results}}
\vspace*{0.5cm}
  \begin{tabular}{cc|ccc|ccc|ccc|cc}
        $\kappa$ & & &
        $m_{\pi}\, a$ & & & $m_{\rho}\, a$ & & &
        $m_{\rm N}\, a$ & & & $m_{\Delta}\, a$      \\ \hline
    0.1278  & & & 0.5400(30) & & & 0.7304(55)  & & & 1.0971(80)  & & & 1.2238(98)  \\
    0.1283  & & & 0.4998(31) & & & 0.7053(58)  & & & 1.0522(84)  & & & 1.1899(102) \\
    0.12885 & & & 0.4521(34) & & & 0.6774(63)  & & & 1,0006(91)  & & & 1.1528(108) \\
    0.1294  & & & 0.3990(38) & & & 0.6491(72)  & & & 0.9465(101) & & & 1.1162(115) \\
    0.1299  & & & 0.3434(43) & & & 0.6228(87)  & & & 0.8944(116) & & & 1.0841(125) \\
    0.13025 & & & 0.2978(47) & & & 0.6040(107) & & & 0.8562(134) & & & 1.0630(135) \\
    0.1306  & & & 0.2419(54) & & & 0.5845(143) & & & 0.8172(171) & & & 1.0443(154) \\
    0.1308  & & & 0.1972(69) & & & 0.5812(213) & & & 0.7950(215) & & & 1.0380(189) \\
\end{tabular}
\end{center}
\end{table}

\section{Octet-Decuplet Mass Splittings}
\label{splitSection}

\begin{figure}[t]
\begin{center}
{\includegraphics[height=0.95\hsize,angle=90]{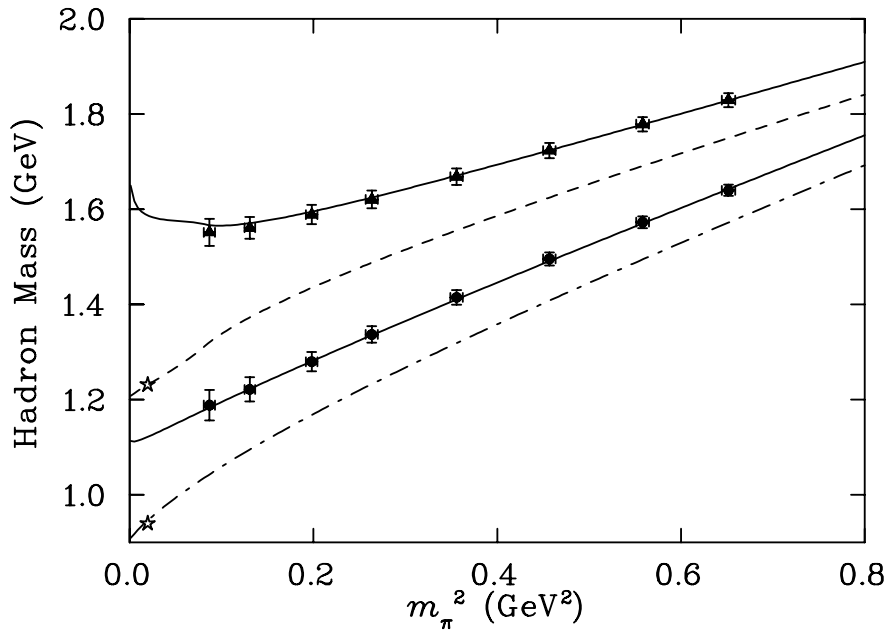}}
\vspace*{0.5cm}
\caption{Nucleon and $\Delta$ masses for the FLIC-fermion action on a
  $20^3 \times 40$ lattice. Here we select $a=0.132$~fm (which
  corresponds to the string tension with $\sqrt{\sigma}=450$~MeV) such
  that the nucleon extrapolation passes through the physical value for
  clarity.  The solid curves illustrate fits of finite-range
  regularised quenched chiral perturbation theory
  \protect\cite{YOUNG,Young:2003ns} to the lattice QCD results.  The
  dashed curves estimate the correction that will arise in unquenching
  the lattice QCD simulations \protect\cite{YOUNG,Young:2003ns}.
  Stars at the physical pion mass denote experimentally measured
  values. }
\label{LQM}
\end{center}
\end{figure}

The results presented in this section are based on an initial
sample of 94 gauge-field configurations of an anticipated 
400 configurations.
Figure~\ref{LQM} shows the $N$ and $\Delta$ masses as a function of
$m_{\pi}^2$ for the FLIC-fermion action on $20^3 \times 40$ lattices
with $a=0.132$~fm (which corresponds to a string tension scale with
$\sqrt{\sigma}=450$~MeV) such that the nucleon extrapolation passes
through the physical value for clarity. An upward curvature in the
$\Delta$ mass for decreasing quark mass is observed in the FLIC
fermion results.  This behaviour, increasing the quenched $N-\Delta$
mass spitting, was predicted by Young {\em et al.}
\cite{YOUNG,Young:2003ns} using 
quenched chiral perturbation theory (Q$\chi$PT) formulated with a
finite-range regulator.  A fit to the FLIC-fermion results is
illustrated by the solid curves.  The dashed curves estimate the
correction that will arise in unquenching the lattice QCD simulations
\cite{YOUNG,Young:2003ns}.
We note that after we have corrected for the absence of sea quark
loops, our results agree simultaneously with the physical values for
both the nucleon and $\Delta$.

\begin{figure}[!]
\begin{center}
{\includegraphics[height=0.95\hsize,angle=90]{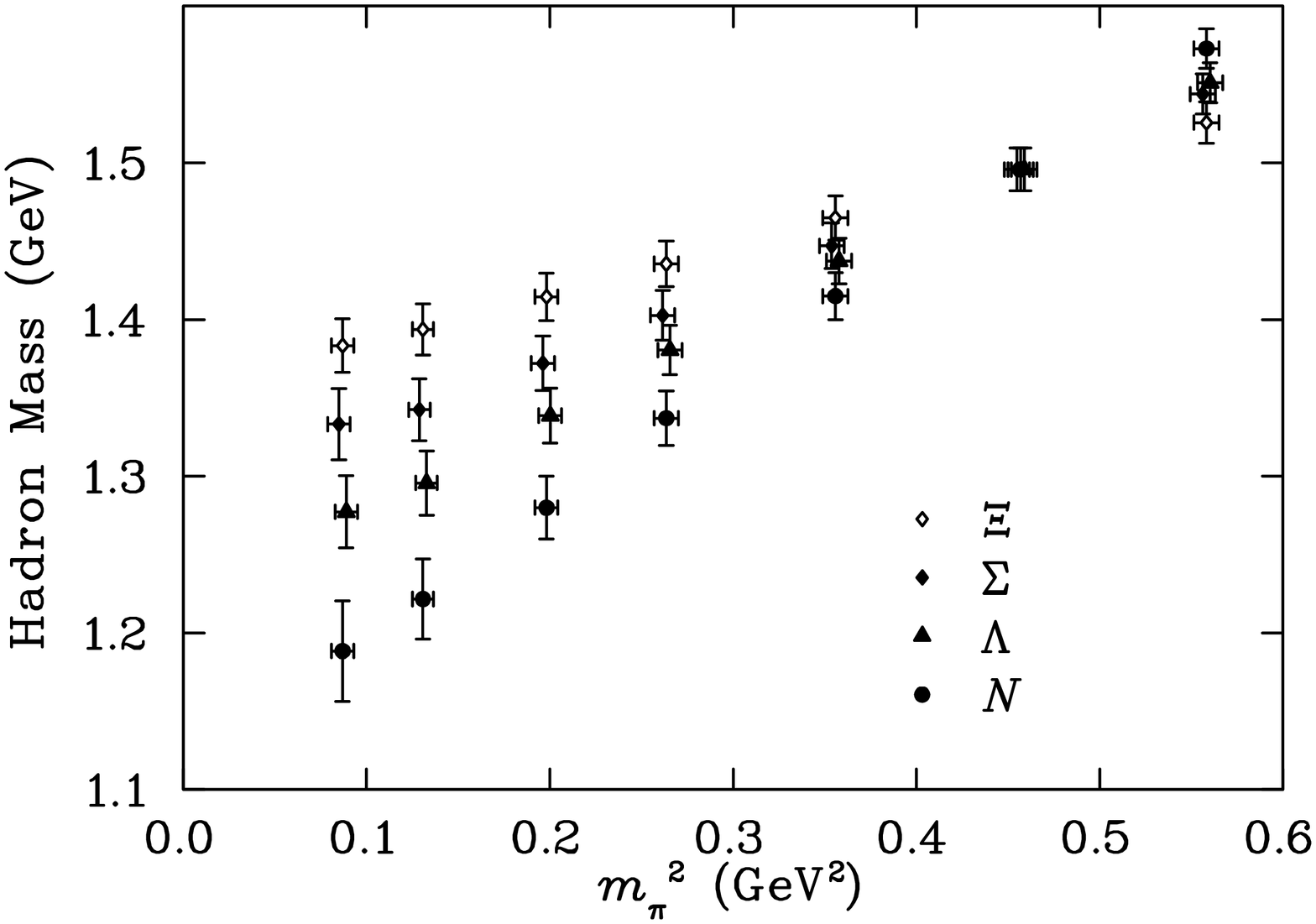}}
%
\vspace*{2cm}
{\includegraphics[height=0.95\hsize,angle=90]{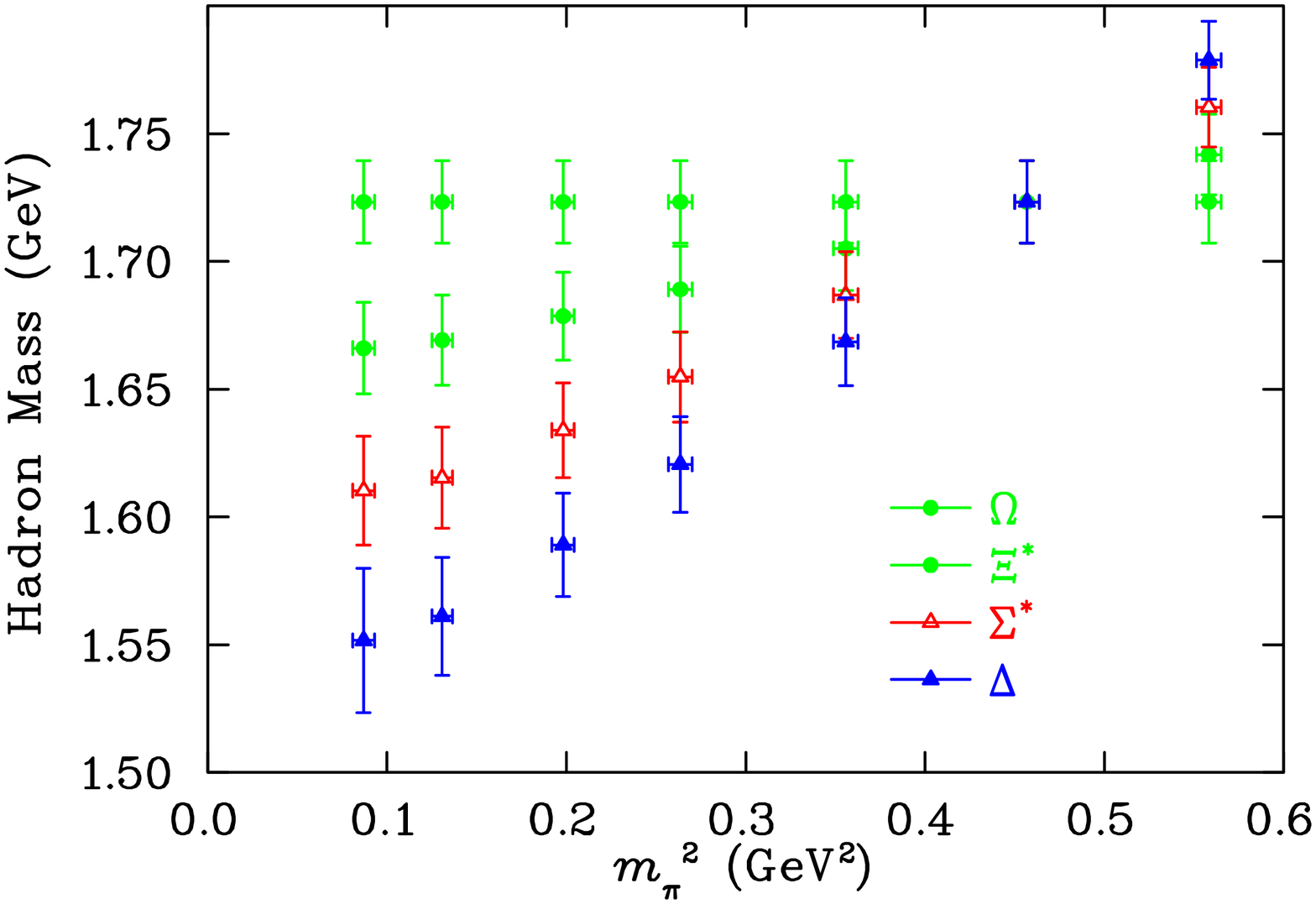}}
\vspace*{0.5cm}
\caption{Octet (top) and decuplet (bottom) baryon masses for the
  FLIC-fermion action on a $20^3 \times 40$ lattice
  with $a=0.134$~fm. }
\label{octet}
\end{center}
\end{figure}

We also calculate the light quark mass behaviour of the octet and
decuplet hyperons. The strange quark mass is chosen in order to
reproduce the physical strange quark mass according to the
phenomenological value of an $s\bar{s}$ pseudoscalar meson,
\be
m_{ss}^2 = 2m_K^2 - m_\pi^2 .
\ee
Upon substitution of the physical masses for the $\pi$ and $K$ mesons,
this corresponds to an $s\bar{s}$ pseudoscalar meson mass of $\sim
0.470$~GeV$^2$ which occurs at our third heaviest quark mass in the
$20^3 \times 40$ lattice analysis.  The results from this calculation
are given in Table~\ref{453HYPresults} and are illustrated in
Fig.~\ref{octet}.  The results show the correct ordering and in
particular, we notice a mass splitting between the $strangeness=-1$
$(I=1)$ $\Sigma$ and $(I=0)$ $\Lambda$ baryons becoming evident in the
light quark mass regime.

\begin{table}[tb]
\begin{center}
\caption{Values of $\kappa$, the octet $\Lambda,\, \Sigma,\, \Xi$ and
        decuplet $\Sigma^*,\, \Xi^*$ masses on a $20^3 \times 40$
        lattice for the FLIC action with 6 sweeps of smearing at
        $\alpha=0.7$.  A string tension analysis provides $a=0.134(2)$
        fm for $\sqrt\sigma=440$~MeV.\label{453HYPresults}}
\vspace*{0.5cm}
  \begin{tabular}{cc|ccc|ccc|ccc|ccc|cc}
        $\kappa$ & & & $m_{\Lambda}\, a$ & & & 
        $m_{\Sigma}\, a$ & & & $m_{\Xi}\, a$ & & & 
        $m_{\Sigma *}\, a$ & & & $m_{\Xi *}\, a$     \\ \hline
    0.1278  & & & 1.0696(84)  & & & 1.0616(83)  & & & 1.0381(87)  
& & & 1.2002(101) & & & 1.1765(104) \\
    0.1283  & & & 1.0376(86)  & & & 1.0328(86)  & & & 1.0206(88)  
& & & 1.1776(104) & & & 1.1652(106) \\
    0.12885 & & & 1.0006(91)  & & & 1.0006(91)  & & & 1,0006(91)  
& & & 1.1528(108) & & & 1.1528(108) \\
    0.1294  & & & 0.9615(97)  & & & 0.9680(97)  & & & 0.9799(94)  
& & & 1.1284(113) & & & 1.1406(110) \\
    0.1299  & & & 0.9235(106) & & & 0.9383(106) & & & 0.9603(98)  
& & & 1.1070(118) & & & 1.1299(113) \\
    0.13025 & & & 0.8955(117) & & & 0.9178(116) & & & 0.9462(102) 
& & & 1.0930(124) & & & 1.1229(115) \\
    0.1306  & & & 0.8667(137) & & & 0.8980(132) & & & 0.9323(109) 
& & & 1.0806(132) & & & 1.1166(118) \\
    0.1308  & & & 0.8544(154) & & & 0.8919(152) & & & 0.9254(114) 
& & & 1.0772(142) & & & 1.1145(120) \\
\end{tabular}
\end{center}
\end{table}

\begin{figure}
\begin{center}
{\includegraphics[angle=90,width=0.95\hsize]{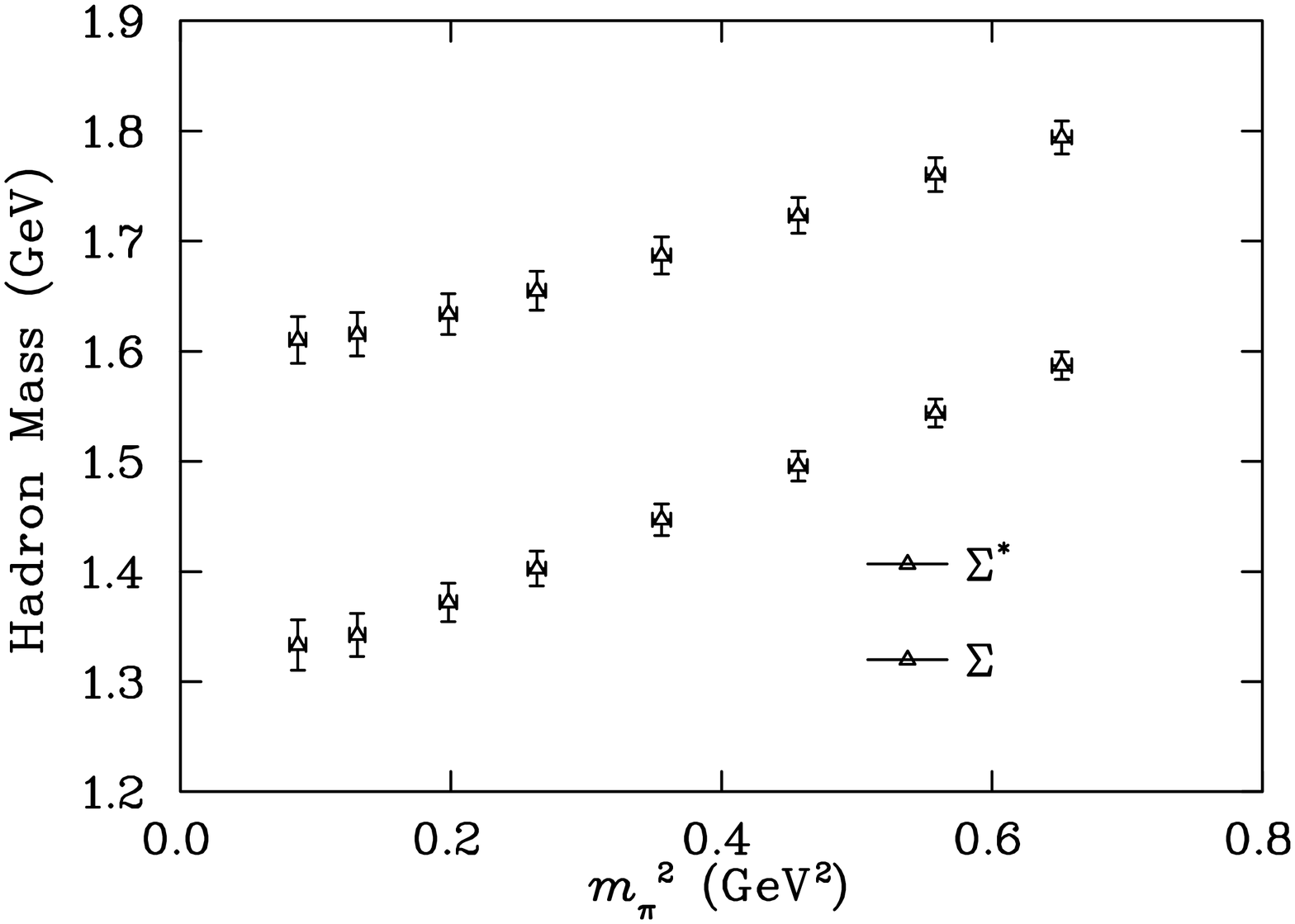}}
%
\vspace*{2cm}
{\includegraphics[angle=90,width=0.95\hsize]{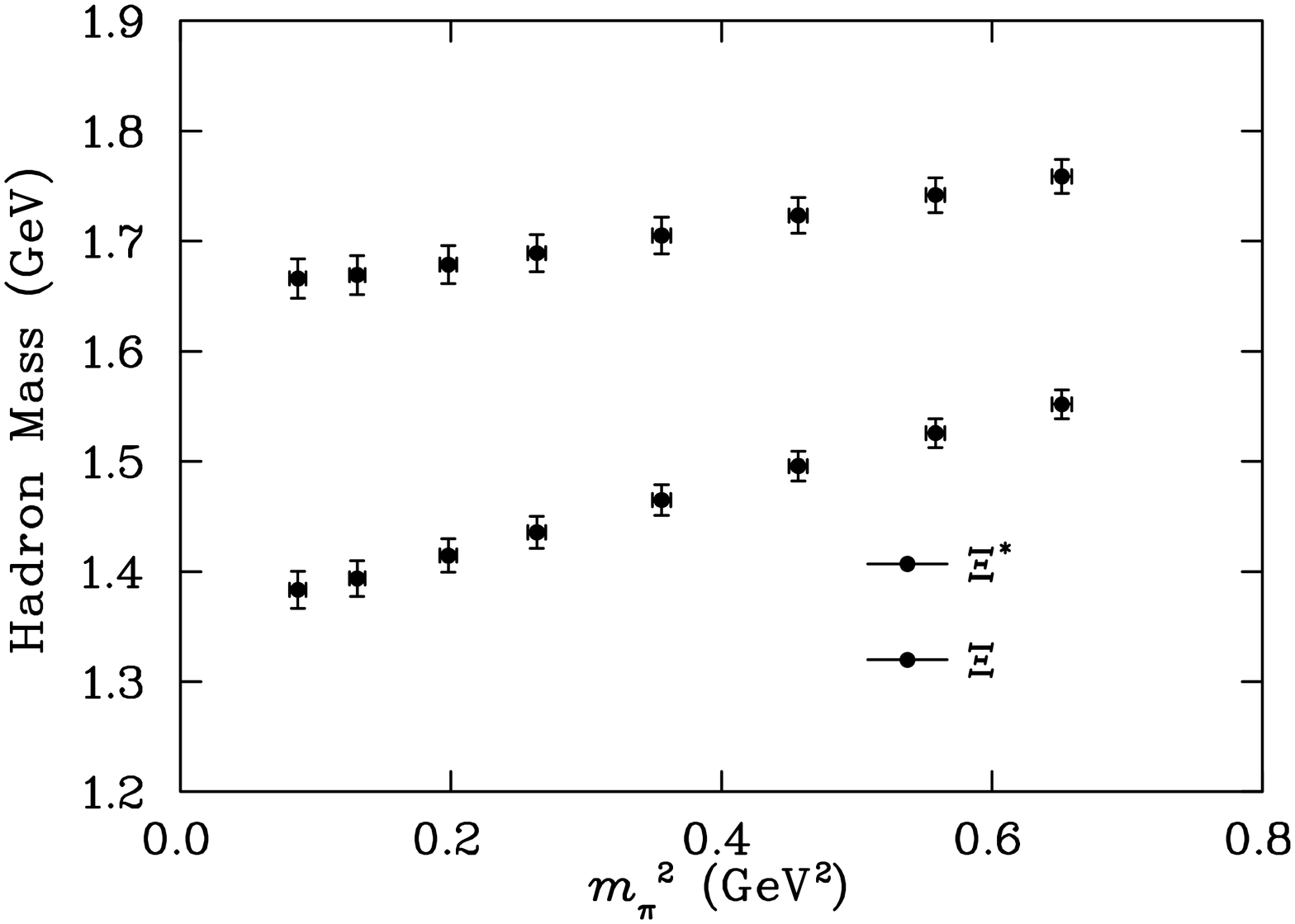}}
\vspace*{0.5cm}
\caption{Octet and decuplet baryon masses for $\Sigma$ (top) and $\Xi$
  (bottom) for the FLIC-fermion action on a $20^3 \times 40$ lattice
  with $a=0.134$~fm. }
\label{XiSplit}
\end{center}
\end{figure}

\begin{figure}[t]
\begin{center}
{\includegraphics[height=0.95\hsize,angle=90]{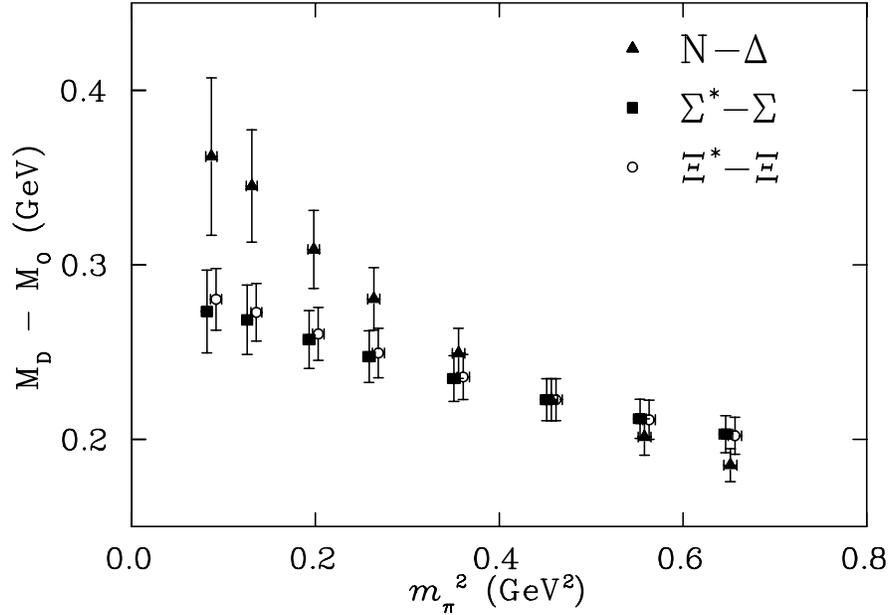}}
\vspace*{0.5cm}
\caption{Decuplet ($M_D$) - octet ($M_O$) baryon mass splittings for
  the FLIC-fermion action on a $20^3 \times 40$ lattice with
  $a=0.132$~fm. }
\label{QDOsplit}
\end{center}
\end{figure}

Just as we saw the non-analytic behaviour of quenched chiral
perturbation theory in the $\Delta$-baryon mass in Fig.~\ref{LQM}
leading to an enhancement of the quenched $N-\Delta$ mass
spitting, Fig.~\ref{XiSplit} shows a similar
enhancement for the decuplet-octet mass splittings in $\Sigma$ and
$\Xi$ baryons respectively. 
The quark model predicts that the hyperfine splittings should
approximately satisfy \cite{Close}
\be
\Xi_s^* - \Xi_s = \mu_s\, \mu_q = \Sigma_s^* - \Sigma_s \, ,
\label{DOsplit}
\ee
where the baryon label denotes the hyperon mass and $\mu_s$ ($\mu_q$)
denotes the magnetic moment of the strange (light) constituent quark.
Fig.~\ref{QDOsplit} shows that even though the quenched approximation
enhances the splitting between octet and decuplet baryons, the
splittings for the $\Sigma$ and $\Xi$ baryons still satisfy
\eq{DOsplit}. 
%
%
%
Agreement of the quenched QCD results with the quark model prediction
is not surprising since both have a suppressed meson cloud.
Similarly, one expects further suppression of the meson cloud when two
(heavy) strange quarks are present in a baryon.

\section{Summary}
\label{FLICconclusion}


We have calculated hadron masses to test the scaling of the Fat-Link
Irrelevant Clover (FLIC) fermion action, in which only the irrelevant,
higher-dimension operators involve smeared links.
One of the main conclusions of this work is that the use of fat links
in the irrelevant operators provides a new form of nonperturbative
${\cal O}(a)$ improvement.
This technique competes well with ${\cal O}(a)$ nonperturbative
improvement on mean field-improved gluon configurations, with the
advantage of a reduced exceptional configuration problem.

Quenched simulations at quark masses down to $m_{\pi}/m_{\rho}=0.35$
have been successfully performed on a $20^3 \times 40$ lattice with a
lattice spacing of 0.134(2)~fm on 94 out of 100 configurations.
Simulations at such light quark masses reveal the non-analytic behaviour
of quenched chiral perturbation theory and provide for an interesting
analysis of the hyperfine splittings between octet and decuplet
baryons.

\section*{Acknowledgements}
We thank Ross Young for contributing the fits of finite-range
regularized quenched chiral perturbation theory to the FLIC fermion
results illustrated in Fig.~\ref{LQM}.  Generous grants of
supercomputer time from the Australian Partnership for Advanced
Computing (APAC) and the Australian National Computing Facility for
Lattice Gauge Theory are gratefully acknowledged.  This work was
supported in part by the Australian Research Council and by DOE
contract DE-AC05-84ER40150 under which the Southeastern Universities
Research Association (SURA) operates the Thomas Jefferson National
Accelerator Facility.

%

\end{document}